\documentclass[aps,prx,twocolumn,showpacs,superscriptaddress,groupedaddress,longbibliography]{revtex4-1}
\usepackage[compat=1.0.0]{tikz-feynman}
\usepackage{amssymb}
\usepackage{bm}
\usepackage{amsmath}
\usepackage{graphicx}
\usepackage{epstopdf}
\usepackage{subfigure}
\usepackage{natbib}
\usepackage{epsfig}
\usepackage{amsfonts}
\usepackage{mathrsfs}
\usepackage[toc,page,title,titletoc,header]{appendix}
\usepackage[colorlinks,linkcolor=blue,citecolor=blue,anchorcolor=blue]{hyperref}
\usepackage{dsfont,amsthm,amsbsy}

\def\be{\begin{equation}}       \def\ee{\end{equation}}
\def\bea{\begin{eqnarray}}      \def\eea{\end{eqnarray}}
\def\ba{\begin{array} }
\def\ea{\end{array} }
\def\bnum{\begin{enumerate} }
\def\enum{\end{enumerate}}

\def\=>{\Rightarrow}
\def\>{\rightarrow}

\def\eye2{Fathbb{I}}

\begin{document}

\title{Solitons with irrational charge in a one-dimensional quantum ferroelectric}

\author{Sijia Zhao}
\affiliation{Department of Applied Physics, Stanford University, Stanford, CA 94305, USA}

\author{Steven~A.~Kivelson}
\affiliation{Department of Physics, Stanford University, Stanford, CA 94305, USA}

\date{\today}

\begin{abstract}
We use analytic (mean-field) and density-matrix renormalization group (DMRG) methods to study a simplified one-dimensional version of the SSH-Holstein model of electron-phonon interactions with one electron per site (half filling) where,  as a function of the ratio of the two couplings, three distinct insulating phases arise - a site-centered charge-density-wave (CDW), a bond-centered bond-density-wave (BDW), and an intermediate ferroelectric (inversion-symmetry breaking) phase with coexisting CDW and BDW order.  In the intermediate phase our DMRG results establish the existence of point-like soliton excitations with irrational charge (that depends on a ratio of coupling constants) and either spin $\frac{1}{2}$ or spin 0, and a surprisingly light effective mass. 
\end{abstract}

\maketitle

\begin{figure}[h!]
    \centering
\includegraphics[width=1.0\linewidth]{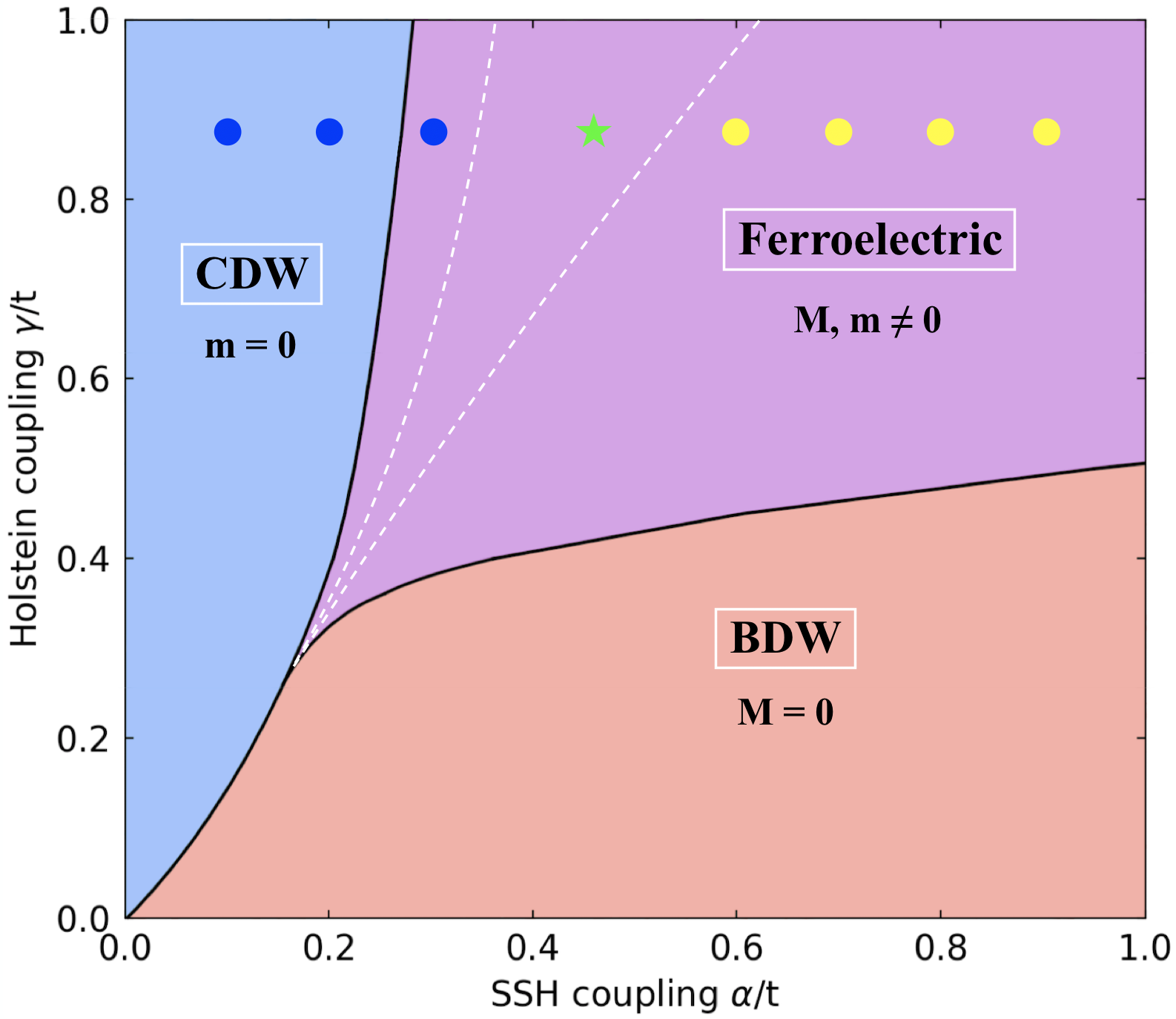}
\caption{\label{fig:phase}\textbf{Mean-field phase diagram} for the present model (Eq.~\eqref{eq:H}) as a function of 
$\alpha/t$ and $\gamma/t$ for fixed quantum parameters, $b_x/t=h_x/t=0.14$.
The dashed lines are a sketch of the actual phase boundary, drawn to be consistent with the DMRG results.
The purple region is the ferroelectric phase where both the SSH ($m$) and the Holstein ($M$) order parameters are non-zero, while in the phases on either side only $M$ or $m$ is non-zero. The dots indicate points at which DMRG calculations have been carried out and the star the particular point in the ferrolectric phase for which the exploration of the soliton dynamics has been carried through. }
\end{figure}

The Su-Schrieffer-Heeger (SSH) model of the electron-phonon coupling\cite{ssh} leads to an insulating broken symmetry (dimerized) state when there is one electron per site (a half-filled band); the associated low energy excitations are topological solitons\cite{Wu2,Wu3,Huda,Go,Kagawa,Horovitz,Wu4,Tokura,Soni,Rice,Muten,Imajo,Kawakami,Marletto,Teemu,Takahashi,Seidel,Ho,Lee,Erez,Dutta,Zhang,Yefsah,Brey,Sondhi,Brazovskii1,Brazovskii2,Takayama,HOROVITZ198061,RICE1979152,Monceau} which exhibit spin-charge separation, i.e. neutral solitons with spin $\frac{1}{2}$ and charge $\pm e$ solitons with spin 0.  
The nature of the quantum number fractionalization involved was further elucidated by Mele and Rice\cite{meleandrice} and Jackiw and Rebbi\cite{jackiw}, who considered the case of dimerization in a generalized version of the SSH model to encompass the case of a binary alloy with alternating A and B type atoms; they showed that the charge of the spinless soliton becomes an irrational number that depends on the difference in the site energy on the two types of atoms.  This was extended\cite{kivelson} to the case of a single-constituent chain but with the electrons coupled both to an acoustic mode (i.e. the SSH coupling) and to an optic mode as in the Holstein model.  At mean-field (MF) level, the phase diagram is roughly that shown in Fig.~\ref{fig:phase}, with
an SSH dominated BDW phase, a Holstein dominated CDW phase, and an intermediate coexistent phase with spontaneously broken inversion symmetry and solitons with irrational charge. 

The present work extends these works in several ways.  Firstly, we identify the state with both CDW and BDW type dimerization as a ferroelectric phase (FE).  We define a model with all the same symmetries in which the phonon modes are represented by pseudo-spins, so that the full quantum problem (beyond mean-field theory) can be efficiently studied using DMRG methods.  We confirm the basic structure of the mean-field phase diagram - shown in Fig.~\ref{fig:phase} -  and in particular the existence of a
range of parameters (albeit one that is considerably narrower than suggested by MF theory) in which the ground state of the half-filled band is a ferroelectric insulator. Focusing on this regime, we find a rich variety of soliton excitations with the expected sorts of fractional quantum numbers and with an exceptionally small effective mass.

{\bf The model:}
We consider the following model of a one-dimensional electron gas coupled to two effective phonon modes:
\begin{align}
\label{eq:H}
H=&-\sum_j \Bigg[t\hat{B}_j +\alpha\tau^z_j\left(\hat B_j-B\right)+b_x\tau_j^x\Bigg]\nonumber \\
&-\sum_j \Bigg[\gamma\sigma^z_j\left(\hat n_j-n\right)+h_x\sigma_j^x\Bigg]
\end{align}
where with $\hat{c}^\dagger_{j,\sigma}$ the fermionic creation operator for an electron with spin-polarization $\sigma$ on site $j$, the bond-charge and site-charge densities are defined as
\begin{align}
\hat{B}_j= \sum_\sigma\left[\hat{c}_{j+1,\sigma}^\dagger \hat{c}_{j,\sigma}+ h.c.\right] \ ; \ \hat n_j
=\sum_\sigma \hat{c}_{j,\sigma}^\dagger \hat{c}_{j,\sigma},
\end{align}
and $B$ and $n$ are, respectively, the average values of $\hat{B}_j$ and $\hat{n}_j$.
The ``SSH'' and ``Holstein'' phonons are represented, respectively, by  pseudo-spin operators $\vec{\tau}_j$ and $\vec{\sigma}_j$ (rather than the  continuous lattice displacement and momentum operators that would appear in a more ``realistic'' version of the model) such that  $\tau_j^z$ and $\sigma_j^z$ represent a phonon displacement at position $j$, and $\tau_j^x$ and $\sigma_j^x$ are the associated phonon kinetic energies.  Thus, $\alpha$ and $\gamma$ play the role of the electron-phonon couplings while $b_x$ and $h_x$ should be interpreted as being the associated phonon frequencies.  The readily solvable classical limit of this model (no phonon dynamics) corresponds to $h_x=b_x=0$.  In the  extreme quantum limit, the phonons can be integrated out to leave a new effective model with  effective bond-charge and site-charge attractions, $W^{\text{eff}}\equiv\alpha^2/b_x$ and $U^{\text{eff}}\equiv\gamma^2/h_x$ for $h_x/t \ \& \ b_x/t \gg 1$.

\begin{figure}[h!]
    \centering
\includegraphics[width=1.0\linewidth]{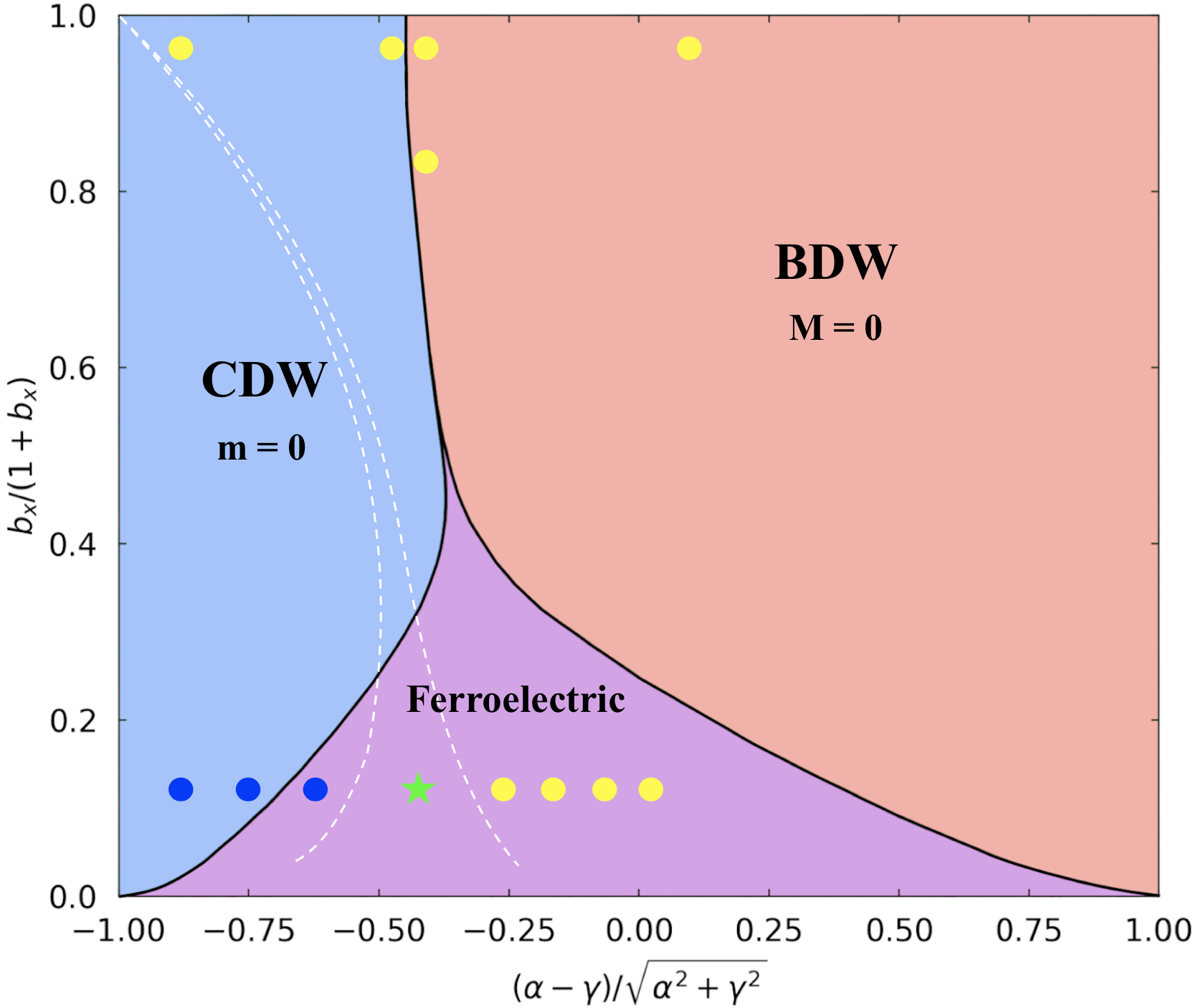}
\caption{\label{fig2:phase}\textbf{Mean-field phase diagram} with symbols as defined in Fig.~\ref{fig:phase}, but here   as a function of the magnitude of the phonon frequency, $b_x/t$, and the relative strength of the two couplings, $(\alpha-\gamma)/\sqrt{\alpha^2+\gamma^2}$, for fixed ratio of the effective phonon frequencies $b_x/h_x= 1$. The solid lines indicate the phase boundaries from the mean-field calculation, which as $b_x$ increases shows a rapidly narrower coexisting region and a negative slope meaning that at fixed coupling ratio there is a re-entrant phase transition.
The colored symbols represent the results of DMRG calculations, where the blue and yellow circles indicate  parameters for which the ground state was found to be pure CDW or BDW, respectively, and the green star the values used in our study of solitons below.}
\end{figure}

{\bf Phase diagram:} 
We start by establishing the phase diagram of the model with one electron per site ($n=1$).  To begin with, treating the problem in mean-field approximation, we find the phase diagram shown in Figs.~\ref{fig:phase} and \ref{fig2:phase}, in which the ground state is always gapped with one of three possible patterns of broken symmetry, CDW, BDW, or coexistence.  It is then relatively straightforward to confirm that the mean-field phase diagram is qualitatively correct in the semiclassical limit, $|h_x/t|$ \& $|b_x/t| \ll 1$.  However, the validity of the mean-field results in the highly quantum (anti-adiabatic) limit, $|h_x/t|$ \& $|b_x/t| \gg 1$, is less obvious.  Firstly, there  is the possibility that one or the other phases might quantum melt,  giving rise to a Luther-Emery (LE) liquid phase (with a spin-gap but no charge gap) that does not appear in the mean field phase diagram at all.  (Indeed, we recently showed\cite{sijia} that this actually does occur in the original Holstein model.)  We present a renormalization group argument that this does not occur in the present model; in particular the LE liquid phase that arises at $h_x/t\to \infty$ in the absence of SSH coupling ($\alpha=0$) is unstable to BDW order in the presence of even weak SSH coupling. None-the-less, these arguments lead us to expect that the position of the phase boundaries differ significantly from the results of mean-field theory in the anti-adiabatic, as sketched in Fig. \ref{fig2:phase} \footnote{At mean-field level, the transition from the CDW to the BDW always occurs through an intermediate FE phase, but this phase gets exponentially small as quantum fluctuations become increasingly strong. It is an open question whether it becomes a direct, fluctuation-driven first-order transition when effects beyond mean-field theory are included.}. Finally, we present DMRG results that corroborate these conclusions.

{\bf a) Mean-Field results:}
To obtain a mean-field phase diagram, we introduce the trial Hamiltonian,
\begin{align}
&H_{\text{tr}}=H_{\text{el}}+H_{\text{pspin}}\\
&H_{\text{el}}=-\sum_j \Bigg[(1+\alpha 
\tau_j)\left(\hat B_j-B\right)+\gamma
\sigma_j\left(\hat n_j-n\right)\Bigg] \nonumber\\
&H_{\text{pspin}}=-\sum_j \Bigg[\alpha
m_j\tau_j^z + b_x\tau_j^x 
+\gamma 
M_j\sigma_j^z + h_x\sigma_j^x\Bigg] \nonumber
\end{align}
where minimizing the corresponding variational energy results in the self-consistency relations
\begin{align}
&m_j=\langle \left(\hat B_j - B \right) \rangle ,  &M_j= \langle\left(\hat n_j - n\right)\rangle  \nonumber\\
&\tau_j=\langle \tau_j^z \rangle ,  &\sigma_j= \langle \sigma_j^z \rangle .
\end{align}
in which the expectation values are with respect to the ground state of $H_{\text{tr}}$.  We look for self-consistent solutions of the form:
$m_j=(-1)^j m$, $M_j=(-1)^j M$, $\tau_j=(-1)^j \bar\tau$, and $\sigma_j=(-1)^j \bar\sigma$. The CDW phase occurs where $\bar\sigma$ \& $M\neq 0$ but $\bar \tau$ \& $m=0$, the BDW phase where $\bar \sigma$ \& $M= 0$ but $\bar\tau$ \& $m\neq 0$, and the ferroelectric phase where all four variational parameters are non-zero.  The explicit calculations that lead to these results are standard, and are summarized in the Supplemental Material.
 
{\bf b) Classical and adiabatic limits:}
It is relatively straightforward to confirm the qualitative validity of the mean-field phase diagram in the adiabatic limit where $|b_x/t|$ \& $|h_x/t|\ll 1$. To begin with, in the classical limit, $b_x=h_x=0$, the pseudo-spins are static variables which in the ground state means $\tau_j^z=(-1)^j$ and $\sigma_j^z=(-1)^j$, i.e. the ground state is ferroelectric 
with coexisting CDW and BDW order. The electronic gap in this state is easily seen to be $2\Delta=4 \sqrt{4\alpha^2+\gamma^2}$; the ferroelectic state is thus stable for a non-zero range of $h_x$ \& $b_x$.    

On the other hand, this range depends not only on $\Delta$, but also on the ratio $\alpha/\gamma$. For instance, for fixed small but non-zero $b_x/t, h_x/t>0$ and $\alpha=0$, the ground state manifestly has $\sigma_j^z=(-1)^j$ and $\tau_j^z = -1$, i.e. it is a CDW, which again because it is fully gapped is perturbatively stable for a range of small but non-zero $\alpha$. Indeed, it is clear even at mean-field level, that so long as  $|b_x/t|$ \& $|h_x/t|\ll 1$ and $\alpha \ll \gamma$,  the phase boundary between the CDW and FE phases occurs where $b_x \sim \alpha^2$.

{\bf c) Anti-adiabatic limit:}
We  can also establish the qualitative validity of the mean-field model in the extreme quantum limit, $|b_x/t|$ \& $ |h_x/t|\gg 1$,  although here the analysis is somewhat more subtle.  In this limit, the pseudo-spins can be integrated out up to leading order in $|t/b_x|$ \& $|t/h_x|$ (second-order perturbation theory) to obtain the effective electronic Hamiltonian 
\begin{align}
H_{\text{eff}}&=-\sum_j\left\{ \hat B_j + \frac{\alpha^2}{2b_x}\left(\hat B_j-B\right)^2 + \frac{\gamma^2}{2h_x}\hat n_j(\hat n_j-1)\right\} 
\nonumber
\end{align}
For $\gamma=0$, $H_{\text{eff}}$ is the same as obtained in the anti-adiabatic limit of the original SSH model. This was analyzed in Ref.\cite{fradkinhirsch} and shown to have a dimerized (BDW) ground state. Because it is a gapped state, it is perturbatively stable for a finite range of non-zero $\gamma$.

The situation is more subtle for $\alpha=0$, where as in the case of the original Holstein model, $H_{\text{eff}}$ is equivalent to the   Hubbard model with $U^{\text{eff}}=-\gamma^2/h_x$, and has a corresponding charge SU(2) symmetry that precludes CDW order, i.e. in this limit the model exhibits a LE liquid phase with a spin-gap and power-law CDW and superconducting correlations (quasi-long-range order).  This phase is not captured by the mean-field treatment.  However, this state has a variety of divergent susceptibilities including a divergent susceptibility to BDW order - as was shown long ago in the context of the spin-$\frac{1}{2}$ Heisenberg antiferromagnet\cite{crossandfisher}.  It thus is likely that this LE phase is unstable to BDW order for any non-zero $\alpha$.  

This analysis leads to the conjecture that in the extreme quantum limit, the system forms a BDW phase for all $\alpha/\gamma$.  This implies that fluctuations shift the boundary between the BDW and CDW phases, as shown in Fig.~\ref{fig2:phase}, but that the topology of the phase diagram, and the lack of a LE phase are correctly captured by MF theory.  This conjecture is consistent with DMRG results as far as we have explored the various parameter ranges.

\begin{figure}[h!]
    \centering
\includegraphics[width=1.0\linewidth]{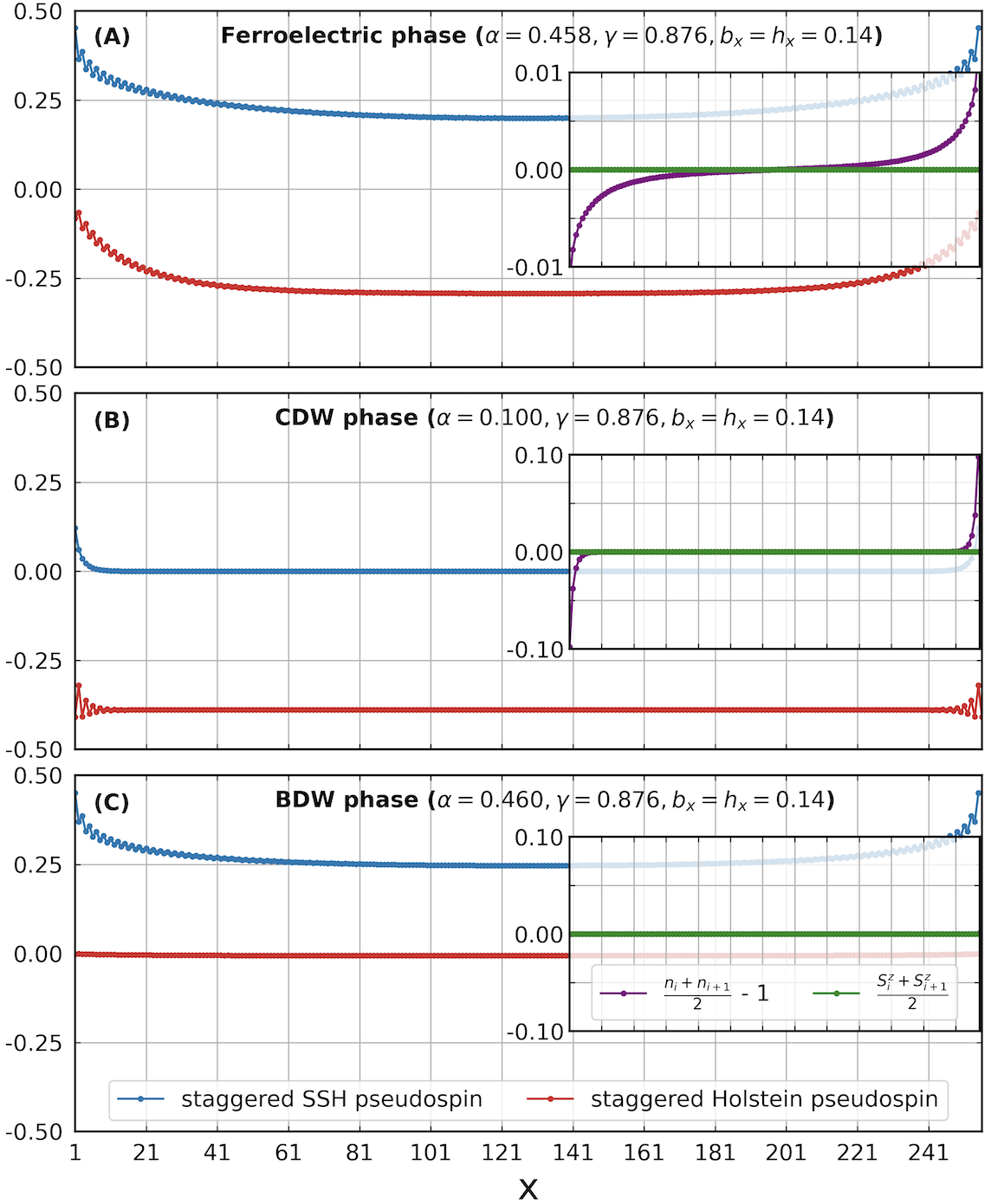}
\caption{\label{fig:3phase}Staggered SSH (red) and Holstein (blue) pseudo-spin order from DMRG for chains of length 256  in (from top to bottom)  the ferroelectric phase, the CDW (Holstein) phase, and the BDW (SSH) phase.  Hamiltonian parameters are given at the top of each panel.  Insets show the smoothed electron density (purple) and spin density (green). 
}
\end{figure}

{\bf d) DMRG results:}
We have also explored the ground-state properties of this Hamiltonian  using high precision DMRG on long (up to length L=260) systems,  
for various choices of parameters $\alpha,\gamma\in[0,1]$ and $b_x, h_x\in[0,50]$, i.e., from the classical to the quantum limit. With relatively small $b_x$ \& $h_x$, a FE phase with both orders coexisting is found, although in a much narrower range of parameters than suggested by mean-field theory.
Results for a set of parameters in the FE phase are shown in Fig.~\ref{fig:3phase}(A), where the red dots represent the staggered Holstein order parameter, $(-1)^j\langle\sigma^z_j\rangle$, the blue dots represent the staggered SSH order parameter, $(-1)^j\langle\tau^z_j\rangle$, and the inset shows the profile of the electron charge and spin densities.  Note that  both order parameters approach a non-vanishing value (the anomalous expectation value) in the bulk and that the system has a  net dipole moment, as reflected in the pile-up of charge density of opposite signs in a broad regime near each of the edges of the system.  

The fragility of the ferroelectric state is illustrated in Figs.~\ref{fig:3phase} (B) and (C), which show the same quantities for parameters that differ only slightly from those in panel (A), but which none-the-less exhibit only one non-zero order parameter and no apparent dipole moment, beyond a small feature at either end of the chain that  reflects an asymmetry of the boundary conditions. The DMRG results have been carried out with bond dimensions up to  $m=5000$, at which point no significant bond-dimension dependence remains, and the typical truncation error is $\epsilon \sim 10^{-10}$.  Further DMRG results and details of the calculations are presented in the Supplemental Material. 

{\bf Solitons in the semiclassical limit:} In this limit (i.e. where the phonon frequencies are small compared to the electronic gap),  the nature of the excitations in both paraelectric phases is understood; for a review, see Ref.\cite{HKSS}.  In both cases, in addition to polaronic excitations with conventional quantum numbers, there exist charge $\pm e$ solitons ($S_e$) with spin $s=0$, and neutral solitons ($S_0$) with spin $\frac{1}{2}$.  Deep in the SSH phase,  where fluctuations of the Holstein mode can be neglected, the creation energy of the neutral and charged solitons are the same due to particle-hole symmetry (i.e. the existence of a mid-gap state). This degeneracy is lifted to linear order in $U^{\text{eff}}$\cite{kivelsonheim}. Effects of quantum fluctuations (finite $h_x$ and/or $b_x$) on these properties merit further exploration\cite{fradkinhirsch,sijia}.

\begin{figure}[b!]
    \centering
\includegraphics[width=1.0\linewidth]{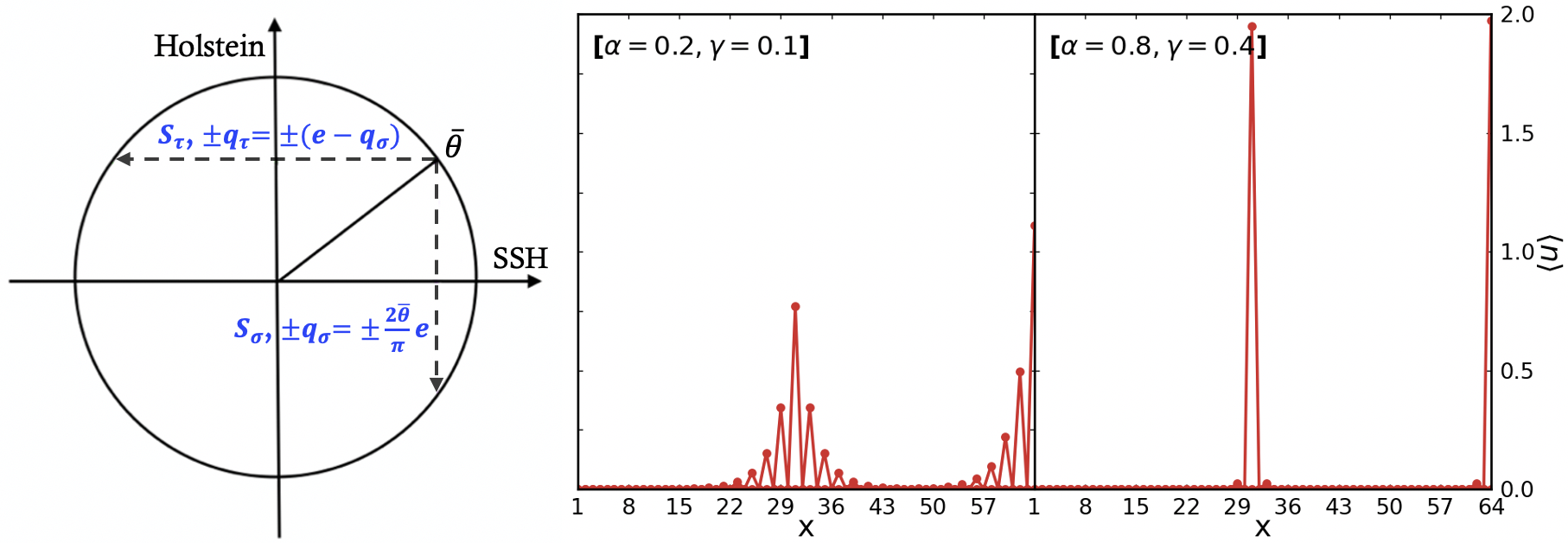}
\caption{\label{fig:theta}\textbf{}Left: schematic of $\bar{\theta}$ for two 
types of solitons. Right: Charge density distribution for an $S_\tau$ located at $R=30$, computed in the classical limit $h_x=b_x=0$. }
\end{figure}
For the present, we focus on the nature of the excitations in the ferroelectric phase. We begin by discussing the solitons as they would emerge 
(in familiar fashion) from a saddle-point approximation to an appropriate Landau-Ginzberg-type effective action. Here, the solitons are spatially localized topological textures in the order parameter fields - that is to say domain walls between regions of distinct ground-state order parameter configurations. It is convenient to introduce a phase and amplitude variable, $\theta_j$ and $A_j$, such that 
\begin{align}
\langle \tau^z_j\rangle = (-1)^j\bar \tau A_j \cos(\theta_j), \ \ \langle \sigma^z_j\rangle = (-1)^j\bar \sigma A_j \sin(\theta_j)
\end{align}
defined in such a way that the four symmetry-related ferroelectric ground states have $A_j=1$ and $\theta_j = \bar\theta$, $-\bar \theta$,  $\pi- \bar\theta$ and $\pi+\bar \theta$ where $\bar\theta$ has a coupling-constant dependent value in the range $0< \bar \theta < \frac{\pi}{2}$ as shown in Fig.~\ref{fig:theta}. Upon approach to the SSH boundary of the ferroelectric phase, $\bar\theta \to 0$, corresponding to pure SSH order, while at the Holstein boundary, $\bar \theta \to \frac{\pi}{2}$.  There are two fundamentally distinct solitons - one ($S_\sigma$) across which the Holstein order parameter changes sign, corresponding to a phase change of $\Delta \theta = \pm 2\bar \theta$ which has a corresponding irrational charge\cite{suschrieffer,goldstonewilczek} $\pm \frac{q_\sigma}{e}\equiv \pm \frac{2\bar\theta}{\pi}$, and the other ($S_\tau$) across which the SSH order parameter changes sign, i.e. $\Delta \theta = \pm (\pi- 2\bar \theta)$ which has charge $\pm q_\tau \equiv \pm(e-q_\sigma)$. 

To see how this works out at the microscopic (lattice) scale, consider the problem in the classical limit, $h_x=b_x=0$, where the ``phonon'' dynamics can be neglected. For instance, a pseudo-spin configuration $\tau^z_j = (-1)^j \ \eta(R-j)$ and $\sigma^z_j=(-1)^j$ corresponds to an SSH soliton ($S_\tau$), where $\eta$ is the Heaviside function.  In the limit $\alpha/\gamma \to 0$, this soliton has vanishing impact on the electronic structure, i.e. $q_\tau \to 0$. Conversely, as $\alpha/\gamma \to \infty$, this is the usual SSH soliton, i.e. $q_\tau \to e$. The bulk gap, $\Delta$, is non-zero for all values of $\alpha/\gamma$, but there is generally a bound-state associated with the soliton that evolves from being just inside the band-edge for $\alpha/\gamma \ll 1$ to the familiar mid-gap position for $\alpha/\gamma \gg 1$.  The charge density distributions for such a soliton for two values of $\alpha/\gamma$  are shown in the right panel of Fig.~\ref{fig:theta} \cite{S0};  note  that the  fractional charge is reflected as a transfer of charge density from the soliton position to an edge-state at the right-hand edge of the system. Because the system is gapped, these features of the classical limit are necessarily preserved in the presence of phonon dynamics, so long as $h_x$ and $b_x$ are sufficiently small that the system remains in the ferroelectric phase.

The irrational charges $q_\tau$ \& $q_\sigma$ are continuous function of parameters in the FE phase, but the sum of $q_\tau+q_\sigma = e$. Composite excitations can, under appropriate circumstances, be constructed as bound-states of solitons and electrons - for example, it is possible to construct solitons with the same quantum numbers as in the paraelectric phase: $S_e$ as a bound state of $S_\sigma+S_\tau$ and $S_0$ by binding an extra spin $\frac{1}{2}$ charge $\pm e$ electron or hole to $S_e$
\footnote{In the case of the neutral composite with the quantum numbers of $S_0$, it is likely that the lowest energy saddle point configuration is actually an amplitude mode, with $\theta_j=$ constant and $A_j$ changing sign across the width of the soliton.}.

In addition, at this level of approximation, each soliton has non-topological properties that derive from the nature of the saddle point solution:  The soliton creation energy, $E_a$, the  width of the soliton, $\xi_a$, and the soliton effective mass, $M_a^\star$, where $a=\tau$ and $\sigma$ in the above.  Each of these quantities varies as one varies parameters within the ferroelectric phase, in ways that can be calculated in the context of the mean-field theory already described.  They generally have singular (critical) behavior as one or the other edge of the phase is approached.  For example, by minimizing the appropriate Landau-Ginzburg energy, one can see that as $\bar\theta \to 0$ upon approach to the boundary with the BDW phase, $\xi_\tau \sim \bar \theta^{-1}$, $E_\tau \sim \bar \theta^2/\xi_{\tau}$, and $M_\tau^\star/M \sim \bar\theta^2/\xi_{\tau}$ where $M$ is an appropriate measure of the soliton mass far from any phase boundaries.

We will not explore these saddle-point notions in detail, since (as discussed below) we can explore them in all their quantum complexity using DMRG.  There are, however, some qualitative points:  On the one hand, since the FE phase we have found is always narrow (much narrower than in the mean-field approximation), one is always close to a phase boundary.  It thus not a surprise that the solitons we find generally appear to have a small creation energy, $E_a/\Delta \ll 1$, a broad  width, $\xi_a \gg 1$, and a light effective mass, $M^\star_a/M \ll 1$.   Conversely, these same properties account for the fact, already discussed, that the most dramatic qualitative effects of quantum fluctuations are to reduce the stability of the FE phase. 

\begin{figure}[h!]
    \centering
\includegraphics[width=1.0\linewidth]{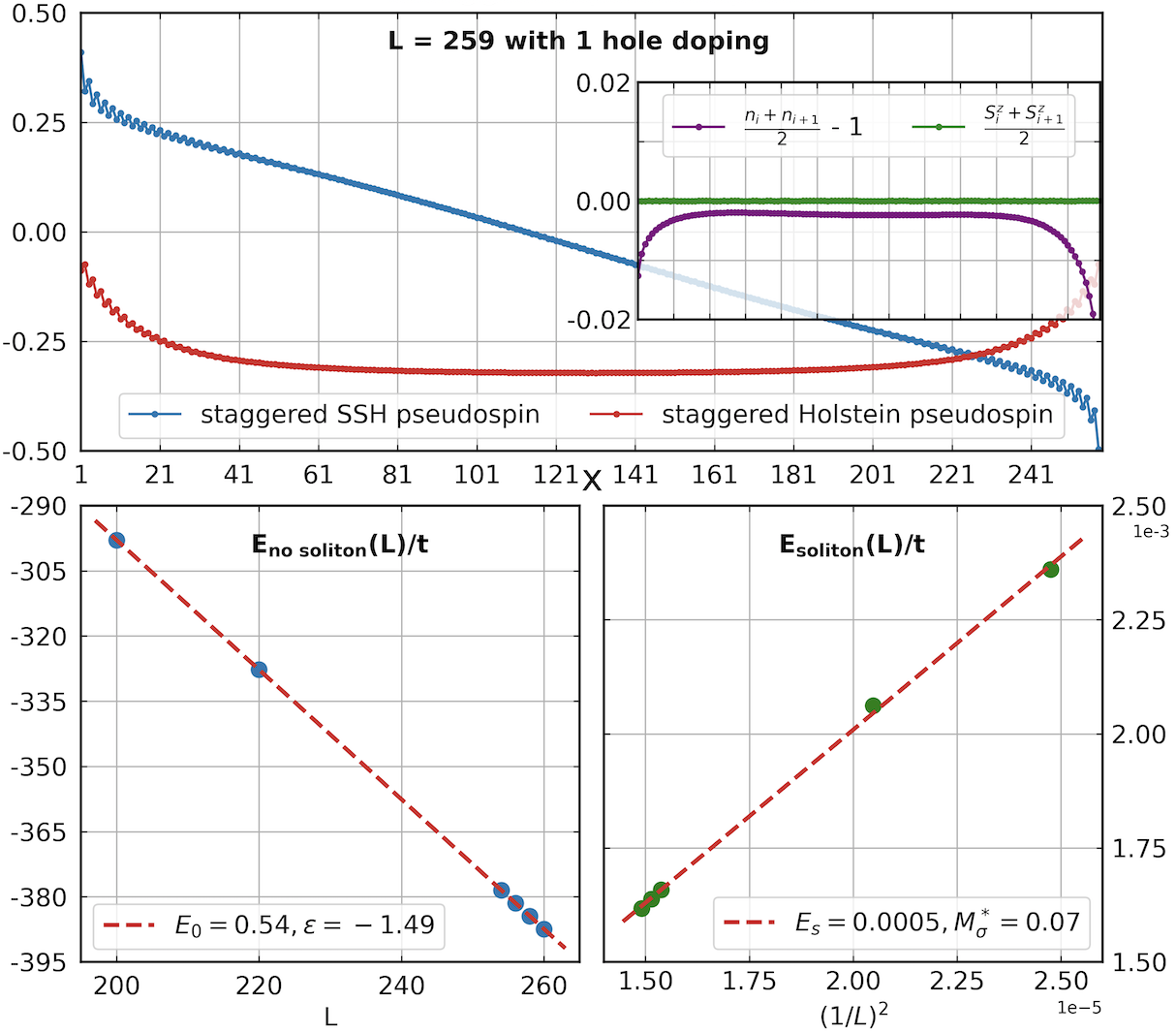}
\caption{\label{fig:oddL-main}\textbf{SSH soliton in an odd-length system with 1 doped hole:} The upper panel shows the staggered pseudo-spin orders and (inset) the smoothed electron and spin densities for $L=259$ and $N=258$.
The lower two panels show the ground-state energies, $E(L,N)$, of the soliton-free system (left panel) and in the presence of a single soliton (right panel).  These, in turn, are used to determine the soliton creation energy and effective mass, as described in the text.}
\end{figure}

{\bf Solitons in DMRG:} We carried out DMRG calculations in various situations in which soliton formation is induced, either by appropriate boundary conditions, adding a small number of charges (doping), and/or working in a sector with non-zero net spin. A particularly simple case is that in which a single soliton ($S_\tau$) is generated by the mismatched boundary conditions for the SSH order parameter in an odd-length system, as was shown in the pure SSH context in Ref.\cite{WPSUpaper}. DMRG results for a chain of length $L=259$, with 258 electrons (1 hole) are shown in the upper panel of Fig.~\ref{fig:oddL-main}. The main curves show  the expectation value of the staggered pseudo-spins and the inset shows the smoothed charge and spin densities as a function of position. We estimate the soliton creation energy, $E_s$, and effective mass, $M_s$, from an analysis of the ground-state energy, $E(L,N)$, where for   $L=N=2n$ even, the ground-state  is soliton-free, while for   $L=N+1=2n+1$ odd, the ground-state has a single SSH soliton, $S_\tau$. We extract the soliton properties from a finite-size analysis according to
\begin{align}
    &E(L=2n,L) =\epsilon L + E_0 + \ldots\\
    &E(L,L-1=2n) =\epsilon L + E_0 + E_s +\left(\frac{\pi^2}{2M_s}\right)\frac{1}{L^2}+\ldots \nonumber
\end{align}
where $E_0$ is an edge contribution, and in the first expression $\ldots = \mathcal{O}(e^{-L/\xi})$, while in the second $\ldots \mathcal{O}(\frac{t}{L^{\alpha>2}})$. For the parameters shown in the figure, this analysis yields $\epsilon = -1.49t$,  $E_0=0.54t$, $E_s=5\times 10^{-4}\ t$ and $M_s=0.07$.
We have also carried out studies in which the soliton is localized by a pinning potential, but interpreting these results is not straightforward since the light mass of soliton makes it necessary to apply  a strong pinning potential if one wishes to localize it significantly.

A full account of various situations in which different soliton-containing ground states were investigated is discussed in the Supplemental Material, and summarized in TABLE~\ref{Table1}.  
Some examples are:  For a neutral odd-length system, $L=N=2n+1$, instead of a single soliton, both $S_\tau$ and $S_\sigma$ now appear. Doping an even length system with a single electron or hole, $L=N\pm 1 =2n$, produces a  polaronic  region with a reduced magnitude of the SSH order parameter, $|\langle \tau_j \rangle|$, together with a Holstein soliton ($S_{\sigma}$). For an undoped, even length system, $L=N=2n$, solitons can also be produced in response to an unequal number of spin up and spin down electrons,  for instance, flipping 1 spin produces a ground state with two pairs of $S_{\tau}$ and $S_{\sigma}$ solitons.  

\newpage
{\bf Acknowledgments:} We acknowledge helpful discussions with Cheng Peng, Hong-Chen Jiang, Ilya Esterlis, Zhaoyu Han and Gene Mele. The DMRG calculations were performed using the Itensor Library\cite{itensor}. Part of the computational work was performed on the Sherlock cluster at Stanford University. The work was funded, in part, by the Department of Energy, Office of Basic Energy Sciences, under Contract No. DE-AC02-76SF00515.

\bibliography{bib}

\begin{thebibliography}{47}%
\makeatletter
\providecommand \@ifxundefined [1]{%
 \@ifx{#1\undefined}
}%
\providecommand \@ifnum [1]{%
 \ifnum #1\expandafter \@firstoftwo
 \else \expandafter \@secondoftwo
 \fi
}%
\providecommand \@ifx [1]{%
 \ifx #1\expandafter \@firstoftwo
 \else \expandafter \@secondoftwo
 \fi
}%
\providecommand \natexlab [1]{#1}%
\providecommand \enquote  [1]{``#1''}%
\providecommand \bibnamefont  [1]{#1}%
\providecommand \bibfnamefont [1]{#1}%
\providecommand \citenamefont [1]{#1}%
\providecommand \href@noop [0]{\@secondoftwo}%
\providecommand \href [0]{\begingroup \@sanitize@url \@href}%
\providecommand \@href[1]{\@@startlink{#1}\@@href}%
\providecommand \@@href[1]{\endgroup#1\@@endlink}%
\providecommand \@sanitize@url [0]{\catcode `\\12\catcode `\$12\catcode `\&12\catcode `\#12\catcode `\^12\catcode `\_12\catcode `\%12\relax}%
\providecommand \@@startlink[1]{}%
\providecommand \@@endlink[0]{}%
\providecommand \url  [0]{\begingroup\@sanitize@url \@url }%
\providecommand \@url [1]{\endgroup\@href {#1}{\urlprefix }}%
\providecommand \urlprefix  [0]{URL }%
\providecommand \Eprint [0]{\href }%
\providecommand \doibase [0]{http://dx.doi.org/}%
\providecommand \selectlanguage [0]{\@gobble}%
\providecommand \bibinfo  [0]{\@secondoftwo}%
\providecommand \bibfield  [0]{\@secondoftwo}%
\providecommand \translation [1]{[#1]}%
\providecommand \BibitemOpen [0]{}%
\providecommand \bibitemStop [0]{}%
\providecommand \bibitemNoStop [0]{.\EOS\space}%
\providecommand \EOS [0]{\spacefactor3000\relax}%
\providecommand \BibitemShut  [1]{\csname bibitem#1\endcsname}%
\let\auto@bib@innerbib\@empty
%</preamble>
\bibitem [{\citenamefont {Fradkin}\ and\ \citenamefont {Hirsch}(1983{\natexlab{a}})}]{ssh}%
  \BibitemOpen
  \bibfield  {author} {\bibinfo {author} {\bibfnamefont {Eduardo}\ \bibnamefont {Fradkin}}\ and\ \bibinfo {author} {\bibfnamefont {Jorge~E.}\ \bibnamefont {Hirsch}},\ }\bibfield  {title} {\enquote {\bibinfo {title} {Phase diagram of one-dimensional electron-phonon systems. i. the su-schrieffer-heeger model},}\ }\href {\doibase 10.1103/PhysRevB.27.1680} {\bibfield  {journal} {\bibinfo  {journal} {Phys. Rev. B}\ }\textbf {\bibinfo {volume} {27}},\ \bibinfo {pages} {1680--1697} (\bibinfo {year} {1983}{\natexlab{a}})}\BibitemShut {NoStop}%
\bibitem [{\citenamefont {Su}\ \emph {et~al.}(1979)\citenamefont {Su}, \citenamefont {Schrieffer},\ and\ \citenamefont {Heeger}}]{Wu2}%
  \BibitemOpen
  \bibfield  {author} {\bibinfo {author} {\bibfnamefont {W.~P.}\ \bibnamefont {Su}}, \bibinfo {author} {\bibfnamefont {J.~R.}\ \bibnamefont {Schrieffer}}, \ and\ \bibinfo {author} {\bibfnamefont {A.~J.}\ \bibnamefont {Heeger}},\ }\bibfield  {title} {\enquote {\bibinfo {title} {Solitons in polyacetylene},}\ }\href {\doibase 10.1103/PhysRevLett.42.1698} {\bibfield  {journal} {\bibinfo  {journal} {Phys. Rev. Lett.}\ }\textbf {\bibinfo {volume} {42}},\ \bibinfo {pages} {1698--1701} (\bibinfo {year} {1979})}\BibitemShut {NoStop}%
\bibitem [{\citenamefont {Su}\ \emph {et~al.}(1980)\citenamefont {Su}, \citenamefont {Schrieffer},\ and\ \citenamefont {Heeger}}]{Wu3}%
  \BibitemOpen
  \bibfield  {author} {\bibinfo {author} {\bibfnamefont {W.~P.}\ \bibnamefont {Su}}, \bibinfo {author} {\bibfnamefont {J.~R.}\ \bibnamefont {Schrieffer}}, \ and\ \bibinfo {author} {\bibfnamefont {A.~J.}\ \bibnamefont {Heeger}},\ }\bibfield  {title} {\enquote {\bibinfo {title} {Soliton excitations in polyacetylene},}\ }\href {\doibase 10.1103/PhysRevB.22.2099} {\bibfield  {journal} {\bibinfo  {journal} {Phys. Rev. B}\ }\textbf {\bibinfo {volume} {22}},\ \bibinfo {pages} {2099--2111} (\bibinfo {year} {1980})}\BibitemShut {NoStop}%
\bibitem [{\citenamefont {Huda}\ \emph {et~al.}(2020)\citenamefont {Huda}, \citenamefont {Kezilebieke}, \citenamefont {Ojanen}, \citenamefont {Drost},\ and\ \citenamefont {Liljeroth}}]{Huda}%
  \BibitemOpen
  \bibfield  {author} {\bibinfo {author} {\bibfnamefont {Md~Nurul}\ \bibnamefont {Huda}}, \bibinfo {author} {\bibfnamefont {Shawulienu}\ \bibnamefont {Kezilebieke}}, \bibinfo {author} {\bibfnamefont {Teemu}\ \bibnamefont {Ojanen}}, \bibinfo {author} {\bibfnamefont {Robert}\ \bibnamefont {Drost}}, \ and\ \bibinfo {author} {\bibfnamefont {Peter}\ \bibnamefont {Liljeroth}},\ }\bibfield  {title} {\enquote {\bibinfo {title} {Tuneable topological domain wall states in engineered atomic chains},}\ }\href {\doibase 10.1038/s41535-020-0219-3} {\bibfield  {journal} {\bibinfo  {journal} {npj Quantum Materials}\ }\textbf {\bibinfo {volume} {5}},\ \bibinfo {pages} {1--5} (\bibinfo {year} {2020})}\BibitemShut {NoStop}%
\bibitem [{\citenamefont {Go}\ \emph {et~al.}(2013)\citenamefont {Go}, \citenamefont {Kang},\ and\ \citenamefont {Han}}]{Go}%
  \BibitemOpen
  \bibfield  {author} {\bibinfo {author} {\bibfnamefont {Gyungchoon}\ \bibnamefont {Go}}, \bibinfo {author} {\bibfnamefont {Kyeong~Tae}\ \bibnamefont {Kang}}, \ and\ \bibinfo {author} {\bibfnamefont {Jung~Hoon}\ \bibnamefont {Han}},\ }\bibfield  {title} {\enquote {\bibinfo {title} {Solitons in one-dimensional three-band model with a central flat band},}\ }\href {\doibase 10.1103/PhysRevB.88.245124} {\bibfield  {journal} {\bibinfo  {journal} {Phys. Rev. B}\ }\textbf {\bibinfo {volume} {88}},\ \bibinfo {pages} {245124} (\bibinfo {year} {2013})}\BibitemShut {NoStop}%
\bibitem [{\citenamefont {Kagawa}\ \emph {et~al.}(2010)\citenamefont {Kagawa}, \citenamefont {Horiuchi}, \citenamefont {Matsui}, \citenamefont {Kumai}, \citenamefont {Onose}, \citenamefont {Hasegawa},\ and\ \citenamefont {Tokura}}]{Kagawa}%
  \BibitemOpen
  \bibfield  {author} {\bibinfo {author} {\bibfnamefont {F.}~\bibnamefont {Kagawa}}, \bibinfo {author} {\bibfnamefont {S.}~\bibnamefont {Horiuchi}}, \bibinfo {author} {\bibfnamefont {H.}~\bibnamefont {Matsui}}, \bibinfo {author} {\bibfnamefont {R.}~\bibnamefont {Kumai}}, \bibinfo {author} {\bibfnamefont {Y.}~\bibnamefont {Onose}}, \bibinfo {author} {\bibfnamefont {T.}~\bibnamefont {Hasegawa}}, \ and\ \bibinfo {author} {\bibfnamefont {Y.}~\bibnamefont {Tokura}},\ }\bibfield  {title} {\enquote {\bibinfo {title} {Electric-field control of solitons in a ferroelectric organic charge-transfer salt},}\ }\href {\doibase 10.1103/PhysRevLett.104.227602} {\bibfield  {journal} {\bibinfo  {journal} {Phys. Rev. Lett.}\ }\textbf {\bibinfo {volume} {104}},\ \bibinfo {pages} {227602} (\bibinfo {year} {2010})}\BibitemShut {NoStop}%
\bibitem [{\citenamefont {Horovitz}\ and\ \citenamefont {Krumhansl}(1984)}]{Horovitz}%
  \BibitemOpen
  \bibfield  {author} {\bibinfo {author} {\bibfnamefont {B.}~\bibnamefont {Horovitz}}\ and\ \bibinfo {author} {\bibfnamefont {J.~A.}\ \bibnamefont {Krumhansl}},\ }\bibfield  {title} {\enquote {\bibinfo {title} {Solitons in the peierls condensate: Phase solitons},}\ }\href {\doibase 10.1103/PhysRevB.29.2109} {\bibfield  {journal} {\bibinfo  {journal} {Phys. Rev. B}\ }\textbf {\bibinfo {volume} {29}},\ \bibinfo {pages} {2109--2124} (\bibinfo {year} {1984})}\BibitemShut {NoStop}%
\bibitem [{\citenamefont {Su}\ and\ \citenamefont {Epstein}(1993)}]{Wu4}%
  \BibitemOpen
  \bibfield  {author} {\bibinfo {author} {\bibfnamefont {W.~P.}\ \bibnamefont {Su}}\ and\ \bibinfo {author} {\bibfnamefont {A.~J.}\ \bibnamefont {Epstein}},\ }\bibfield  {title} {\enquote {\bibinfo {title} {Optical and magnetic signatures of localized excitations in pernigraniline: Role of neutral solitons},}\ }\href {\doibase 10.1103/PhysRevLett.70.1497} {\bibfield  {journal} {\bibinfo  {journal} {Phys. Rev. Lett.}\ }\textbf {\bibinfo {volume} {70}},\ \bibinfo {pages} {1497--1500} (\bibinfo {year} {1993})}\BibitemShut {NoStop}%
\bibitem [{\citenamefont {Tokura}\ \emph {et~al.}(1989)\citenamefont {Tokura}, \citenamefont {Koshihara}, \citenamefont {Iwasa}, \citenamefont {Okamoto}, \citenamefont {Komatsu}, \citenamefont {Koda}, \citenamefont {Iwasawa},\ and\ \citenamefont {Saito}}]{Tokura}%
  \BibitemOpen
  \bibfield  {author} {\bibinfo {author} {\bibfnamefont {Y.}~\bibnamefont {Tokura}}, \bibinfo {author} {\bibfnamefont {S.}~\bibnamefont {Koshihara}}, \bibinfo {author} {\bibfnamefont {Y.}~\bibnamefont {Iwasa}}, \bibinfo {author} {\bibfnamefont {H.}~\bibnamefont {Okamoto}}, \bibinfo {author} {\bibfnamefont {T.}~\bibnamefont {Komatsu}}, \bibinfo {author} {\bibfnamefont {T.}~\bibnamefont {Koda}}, \bibinfo {author} {\bibfnamefont {N.}~\bibnamefont {Iwasawa}}, \ and\ \bibinfo {author} {\bibfnamefont {G.}~\bibnamefont {Saito}},\ }\bibfield  {title} {\enquote {\bibinfo {title} {Domain-wall dynamics in organic charge-transfer compounds with one-dimensional ferroelectricity},}\ }\href {\doibase 10.1103/PhysRevLett.63.2405} {\bibfield  {journal} {\bibinfo  {journal} {Phys. Rev. Lett.}\ }\textbf {\bibinfo {volume} {63}},\ \bibinfo {pages} {2405--2408} (\bibinfo {year} {1989})}\BibitemShut {NoStop}%
\bibitem [{\citenamefont {Soni}\ and\ \citenamefont {Baskaran}(1984)}]{Soni}%
  \BibitemOpen
  \bibfield  {author} {\bibinfo {author} {\bibfnamefont {Vikram}\ \bibnamefont {Soni}}\ and\ \bibinfo {author} {\bibfnamefont {G.}~\bibnamefont {Baskaran}},\ }\bibfield  {title} {\enquote {\bibinfo {title} {Depletion of fractional fermion number of a soliton at finite chemical potential and temperature},}\ }\href {\doibase 10.1103/PhysRevLett.53.523} {\bibfield  {journal} {\bibinfo  {journal} {Phys. Rev. Lett.}\ }\textbf {\bibinfo {volume} {53}},\ \bibinfo {pages} {523--526} (\bibinfo {year} {1984})}\BibitemShut {NoStop}%
\bibitem [{\citenamefont {Rice}\ \emph {et~al.}(1983)\citenamefont {Rice}, \citenamefont {Bishop},\ and\ \citenamefont {Campbell}}]{Rice}%
  \BibitemOpen
  \bibfield  {author} {\bibinfo {author} {\bibfnamefont {M.~J.}\ \bibnamefont {Rice}}, \bibinfo {author} {\bibfnamefont {A.~R.}\ \bibnamefont {Bishop}}, \ and\ \bibinfo {author} {\bibfnamefont {D.~K.}\ \bibnamefont {Campbell}},\ }\bibfield  {title} {\enquote {\bibinfo {title} {Unusual soliton properties of the infinite polyyne chain},}\ }\href {\doibase 10.1103/PhysRevLett.51.2136} {\bibfield  {journal} {\bibinfo  {journal} {Phys. Rev. Lett.}\ }\textbf {\bibinfo {volume} {51}},\ \bibinfo {pages} {2136--2139} (\bibinfo {year} {1983})}\BibitemShut {NoStop}%
\bibitem [{\citenamefont {Muten}\ \emph {et~al.}(2024)\citenamefont {Muten}, \citenamefont {Frankland},\ and\ \citenamefont {McCann}}]{Muten}%
  \BibitemOpen
  \bibfield  {author} {\bibinfo {author} {\bibfnamefont {James~H.}\ \bibnamefont {Muten}}, \bibinfo {author} {\bibfnamefont {Louise~H.}\ \bibnamefont {Frankland}}, \ and\ \bibinfo {author} {\bibfnamefont {Edward}\ \bibnamefont {McCann}},\ }\bibfield  {title} {\enquote {\bibinfo {title} {Solitons in binary compounds with stacked two-dimensional honeycomb lattices},}\ }\href {\doibase 10.1103/PhysRevB.109.165416} {\bibfield  {journal} {\bibinfo  {journal} {Phys. Rev. B}\ }\textbf {\bibinfo {volume} {109}},\ \bibinfo {pages} {165416} (\bibinfo {year} {2024})}\BibitemShut {NoStop}%
\bibitem [{\citenamefont {Imajo}\ \emph {et~al.}(2024)\citenamefont {Imajo}, \citenamefont {Miyake}, \citenamefont {Kurihara}, \citenamefont {Tokunaga}, \citenamefont {Kindo}, \citenamefont {Horiuchi},\ and\ \citenamefont {Kagawa}}]{Imajo}%
  \BibitemOpen
  \bibfield  {author} {\bibinfo {author} {\bibfnamefont {Shusaku}\ \bibnamefont {Imajo}}, \bibinfo {author} {\bibfnamefont {Atsushi}\ \bibnamefont {Miyake}}, \bibinfo {author} {\bibfnamefont {Ryosuke}\ \bibnamefont {Kurihara}}, \bibinfo {author} {\bibfnamefont {Masashi}\ \bibnamefont {Tokunaga}}, \bibinfo {author} {\bibfnamefont {Koichi}\ \bibnamefont {Kindo}}, \bibinfo {author} {\bibfnamefont {Sachio}\ \bibnamefont {Horiuchi}}, \ and\ \bibinfo {author} {\bibfnamefont {Fumitaka}\ \bibnamefont {Kagawa}},\ }\bibfield  {title} {\enquote {\bibinfo {title} {Quantum liquid states of spin solitons in a ferroelectric spin-peierls state},}\ }\href {\doibase 10.1103/PhysRevLett.132.096601} {\bibfield  {journal} {\bibinfo  {journal} {Phys. Rev. Lett.}\ }\textbf {\bibinfo {volume} {132}},\ \bibinfo {pages} {096601} (\bibinfo {year} {2024})}\BibitemShut {NoStop}%
\bibitem [{\citenamefont {Kawakami}\ \emph {et~al.}(2023)\citenamefont {Kawakami}, \citenamefont {Tamaki},\ and\ \citenamefont {Koshino}}]{Kawakami}%
  \BibitemOpen
  \bibfield  {author} {\bibinfo {author} {\bibfnamefont {Takuto}\ \bibnamefont {Kawakami}}, \bibinfo {author} {\bibfnamefont {Gen}\ \bibnamefont {Tamaki}}, \ and\ \bibinfo {author} {\bibfnamefont {Mikito}\ \bibnamefont {Koshino}},\ }\bibfield  {title} {\enquote {\bibinfo {title} {Topological domain walls in graphene nanoribbons with carrier doping},}\ }\href {\doibase 10.1103/PhysRevB.108.045401} {\bibfield  {journal} {\bibinfo  {journal} {Phys. Rev. B}\ }\textbf {\bibinfo {volume} {108}},\ \bibinfo {pages} {045401} (\bibinfo {year} {2023})}\BibitemShut {NoStop}%
\bibitem [{\citenamefont {Marletto}\ and\ \citenamefont {Rasetti}(2012)}]{Marletto}%
  \BibitemOpen
  \bibfield  {author} {\bibinfo {author} {\bibfnamefont {Chiara}\ \bibnamefont {Marletto}}\ and\ \bibinfo {author} {\bibfnamefont {Mario}\ \bibnamefont {Rasetti}},\ }\bibfield  {title} {\enquote {\bibinfo {title} {Peierls distortion and quantum solitons},}\ }\href {\doibase 10.1103/PhysRevLett.109.126405} {\bibfield  {journal} {\bibinfo  {journal} {Phys. Rev. Lett.}\ }\textbf {\bibinfo {volume} {109}},\ \bibinfo {pages} {126405} (\bibinfo {year} {2012})}\BibitemShut {NoStop}%
\bibitem [{\citenamefont {V\"ayrynen}\ and\ \citenamefont {Ojanen}(2011)}]{Teemu}%
  \BibitemOpen
  \bibfield  {author} {\bibinfo {author} {\bibfnamefont {Jukka~I.}\ \bibnamefont {V\"ayrynen}}\ and\ \bibinfo {author} {\bibfnamefont {Teemu}\ \bibnamefont {Ojanen}},\ }\bibfield  {title} {\enquote {\bibinfo {title} {Chiral topological phases and fractional domain wall excitations in one-dimensional chains and wires},}\ }\href {\doibase 10.1103/PhysRevLett.107.166804} {\bibfield  {journal} {\bibinfo  {journal} {Phys. Rev. Lett.}\ }\textbf {\bibinfo {volume} {107}},\ \bibinfo {pages} {166804} (\bibinfo {year} {2011})}\BibitemShut {NoStop}%
\bibitem [{\citenamefont {Takahashi}\ and\ \citenamefont {Nitta}(2013)}]{Takahashi}%
  \BibitemOpen
  \bibfield  {author} {\bibinfo {author} {\bibfnamefont {Daisuke~A.}\ \bibnamefont {Takahashi}}\ and\ \bibinfo {author} {\bibfnamefont {Muneto}\ \bibnamefont {Nitta}},\ }\bibfield  {title} {\enquote {\bibinfo {title} {Self-consistent multiple complex-kink solutions in bogoliubov--de gennes and chiral gross-neveu systems},}\ }\href {\doibase 10.1103/PhysRevLett.110.131601} {\bibfield  {journal} {\bibinfo  {journal} {Phys. Rev. Lett.}\ }\textbf {\bibinfo {volume} {110}},\ \bibinfo {pages} {131601} (\bibinfo {year} {2013})}\BibitemShut {NoStop}%
\bibitem [{\citenamefont {Seidel}\ and\ \citenamefont {Lee}(2006)}]{Seidel}%
  \BibitemOpen
  \bibfield  {author} {\bibinfo {author} {\bibfnamefont {Alexander}\ \bibnamefont {Seidel}}\ and\ \bibinfo {author} {\bibfnamefont {Dung-Hai}\ \bibnamefont {Lee}},\ }\bibfield  {title} {\enquote {\bibinfo {title} {Abelian and non-abelian hall liquids and charge-density wave: Quantum number fractionalization in one and two dimensions},}\ }\href {\doibase 10.1103/PhysRevLett.97.056804} {\bibfield  {journal} {\bibinfo  {journal} {Phys. Rev. Lett.}\ }\textbf {\bibinfo {volume} {97}},\ \bibinfo {pages} {056804} (\bibinfo {year} {2006})}\BibitemShut {NoStop}%
\bibitem [{\citenamefont {Ho}\ \emph {et~al.}(1984)\citenamefont {Ho}, \citenamefont {Fulco}, \citenamefont {Schrieffer},\ and\ \citenamefont {Wilczek}}]{Ho}%
  \BibitemOpen
  \bibfield  {author} {\bibinfo {author} {\bibfnamefont {T.~L.}\ \bibnamefont {Ho}}, \bibinfo {author} {\bibfnamefont {J.~R.}\ \bibnamefont {Fulco}}, \bibinfo {author} {\bibfnamefont {J.~R.}\ \bibnamefont {Schrieffer}}, \ and\ \bibinfo {author} {\bibfnamefont {F.}~\bibnamefont {Wilczek}},\ }\bibfield  {title} {\enquote {\bibinfo {title} {Solitons in superfluid $^{3}\mathrm{He}$-$a$: Bound states on domain walls},}\ }\href {\doibase 10.1103/PhysRevLett.52.1524} {\bibfield  {journal} {\bibinfo  {journal} {Phys. Rev. Lett.}\ }\textbf {\bibinfo {volume} {52}},\ \bibinfo {pages} {1524--1527} (\bibinfo {year} {1984})}\BibitemShut {NoStop}%
\bibitem [{\citenamefont {Lee}\ \emph {et~al.}(2019)\citenamefont {Lee}, \citenamefont {Shim}, \citenamefont {Hyun},\ and\ \citenamefont {Kim}}]{Lee}%
  \BibitemOpen
  \bibfield  {author} {\bibinfo {author} {\bibfnamefont {Geunseop}\ \bibnamefont {Lee}}, \bibinfo {author} {\bibfnamefont {Hyungjoon}\ \bibnamefont {Shim}}, \bibinfo {author} {\bibfnamefont {Jung-Min}\ \bibnamefont {Hyun}}, \ and\ \bibinfo {author} {\bibfnamefont {Hanchul}\ \bibnamefont {Kim}},\ }\bibfield  {title} {\enquote {\bibinfo {title} {Intertwined solitons and impurities in a quasi-one-dimensional charge-density-wave system: $\mathrm{In}/\mathrm{Si}(111)$},}\ }\href {\doibase 10.1103/PhysRevLett.122.016102} {\bibfield  {journal} {\bibinfo  {journal} {Phys. Rev. Lett.}\ }\textbf {\bibinfo {volume} {122}},\ \bibinfo {pages} {016102} (\bibinfo {year} {2019})}\BibitemShut {NoStop}%
\bibitem [{\citenamefont {Efroni}\ \emph {et~al.}(2017)\citenamefont {Efroni}, \citenamefont {Ilani},\ and\ \citenamefont {Berg}}]{Erez}%
  \BibitemOpen
  \bibfield  {author} {\bibinfo {author} {\bibfnamefont {Yonathan}\ \bibnamefont {Efroni}}, \bibinfo {author} {\bibfnamefont {Shahal}\ \bibnamefont {Ilani}}, \ and\ \bibinfo {author} {\bibfnamefont {Erez}\ \bibnamefont {Berg}},\ }\bibfield  {title} {\enquote {\bibinfo {title} {Topological transitions and fractional charges induced by strain and a magnetic field in carbon nanotubes},}\ }\href {\doibase 10.1103/PhysRevLett.119.147704} {\bibfield  {journal} {\bibinfo  {journal} {Phys. Rev. Lett.}\ }\textbf {\bibinfo {volume} {119}},\ \bibinfo {pages} {147704} (\bibinfo {year} {2017})}\BibitemShut {NoStop}%
\bibitem [{\citenamefont {Dutta}\ and\ \citenamefont {Mueller}(2017)}]{Dutta}%
  \BibitemOpen
  \bibfield  {author} {\bibinfo {author} {\bibfnamefont {Shovan}\ \bibnamefont {Dutta}}\ and\ \bibinfo {author} {\bibfnamefont {Erich~J.}\ \bibnamefont {Mueller}},\ }\bibfield  {title} {\enquote {\bibinfo {title} {Collective modes of a soliton train in a fermi superfluid},}\ }\href {\doibase 10.1103/PhysRevLett.118.260402} {\bibfield  {journal} {\bibinfo  {journal} {Phys. Rev. Lett.}\ }\textbf {\bibinfo {volume} {118}},\ \bibinfo {pages} {260402} (\bibinfo {year} {2017})}\BibitemShut {NoStop}%
\bibitem [{\citenamefont {Zhang}\ \emph {et~al.}(2011)\citenamefont {Zhang}, \citenamefont {Choi}, \citenamefont {Xu}, \citenamefont {Wang}, \citenamefont {Zhai}, \citenamefont {Wang}, \citenamefont {Zeng}, \citenamefont {Cho}, \citenamefont {Zhang},\ and\ \citenamefont {Hou}}]{Zhang}%
  \BibitemOpen
  \bibfield  {author} {\bibinfo {author} {\bibfnamefont {Hui}\ \bibnamefont {Zhang}}, \bibinfo {author} {\bibfnamefont {Jin-Ho}\ \bibnamefont {Choi}}, \bibinfo {author} {\bibfnamefont {Yang}\ \bibnamefont {Xu}}, \bibinfo {author} {\bibfnamefont {Xiuxia}\ \bibnamefont {Wang}}, \bibinfo {author} {\bibfnamefont {Xiaofang}\ \bibnamefont {Zhai}}, \bibinfo {author} {\bibfnamefont {Bing}\ \bibnamefont {Wang}}, \bibinfo {author} {\bibfnamefont {Changgan}\ \bibnamefont {Zeng}}, \bibinfo {author} {\bibfnamefont {Jun-Hyung}\ \bibnamefont {Cho}}, \bibinfo {author} {\bibfnamefont {Zhenyu}\ \bibnamefont {Zhang}}, \ and\ \bibinfo {author} {\bibfnamefont {J.~G.}\ \bibnamefont {Hou}},\ }\bibfield  {title} {\enquote {\bibinfo {title} {Atomic structure, energetics, and dynamics of topological solitons in indium chains on si(111) surfaces},}\ }\href {\doibase 10.1103/PhysRevLett.106.026801} {\bibfield  {journal} {\bibinfo  {journal} {Phys. Rev. Lett.}\ }\textbf {\bibinfo {volume} {106}},\ \bibinfo {pages} {026801} (\bibinfo
  {year} {2011})}\BibitemShut {NoStop}%
\bibitem [{\citenamefont {Yefsah}\ \emph {et~al.}(2013)\citenamefont {Yefsah}, \citenamefont {Sommer}, \citenamefont {Ku}, \citenamefont {Cheuk}, \citenamefont {Ji}, \citenamefont {Bakr},\ and\ \citenamefont {Zwierlein}}]{Yefsah}%
  \BibitemOpen
  \bibfield  {author} {\bibinfo {author} {\bibfnamefont {Tarik}\ \bibnamefont {Yefsah}}, \bibinfo {author} {\bibfnamefont {Ariel~T.}\ \bibnamefont {Sommer}}, \bibinfo {author} {\bibfnamefont {Mark J.~H.}\ \bibnamefont {Ku}}, \bibinfo {author} {\bibfnamefont {Lawrence~W.}\ \bibnamefont {Cheuk}}, \bibinfo {author} {\bibfnamefont {Wenjie}\ \bibnamefont {Ji}}, \bibinfo {author} {\bibfnamefont {Waseem~S.}\ \bibnamefont {Bakr}}, \ and\ \bibinfo {author} {\bibfnamefont {Martin~W.}\ \bibnamefont {Zwierlein}},\ }\bibfield  {title} {\enquote {\bibinfo {title} {Heavy solitons in a fermionic superfluid},}\ }\href {\doibase 10.1038/nature12338} {\bibfield  {journal} {\bibinfo  {journal} {Nature}\ }\textbf {\bibinfo {volume} {499}},\ \bibinfo {pages} {426--430} (\bibinfo {year} {2013})}\BibitemShut {NoStop}%
\bibitem [{\citenamefont {Brey}\ and\ \citenamefont {Littlewood}(2005)}]{Brey}%
  \BibitemOpen
  \bibfield  {author} {\bibinfo {author} {\bibfnamefont {Luis}\ \bibnamefont {Brey}}\ and\ \bibinfo {author} {\bibfnamefont {P.~B.}\ \bibnamefont {Littlewood}},\ }\bibfield  {title} {\enquote {\bibinfo {title} {Solitonic phase in manganites},}\ }\href {\doibase 10.1103/PhysRevLett.95.117205} {\bibfield  {journal} {\bibinfo  {journal} {Phys. Rev. Lett.}\ }\textbf {\bibinfo {volume} {95}},\ \bibinfo {pages} {117205} (\bibinfo {year} {2005})}\BibitemShut {NoStop}%
\bibitem [{\citenamefont {Moessner}\ and\ \citenamefont {Sondhi}(2010)}]{Sondhi}%
  \BibitemOpen
  \bibfield  {author} {\bibinfo {author} {\bibfnamefont {R.}~\bibnamefont {Moessner}}\ and\ \bibinfo {author} {\bibfnamefont {S.~L.}\ \bibnamefont {Sondhi}},\ }\bibfield  {title} {\enquote {\bibinfo {title} {Irrational charge from topological order},}\ }\href {\doibase 10.1103/PhysRevLett.105.166401} {\bibfield  {journal} {\bibinfo  {journal} {Phys. Rev. Lett.}\ }\textbf {\bibinfo {volume} {105}},\ \bibinfo {pages} {166401} (\bibinfo {year} {2010})}\BibitemShut {NoStop}%
\bibitem [{\citenamefont {Brazovskii}(1978)}]{Brazovskii1}%
  \BibitemOpen
  \bibfield  {author} {\bibinfo {author} {\bibfnamefont {S.~A.}\ \bibnamefont {Brazovskii}},\ }\bibfield  {title} {\enquote {\bibinfo {title} {Electronic excitations in the peierls-fr\"{o}hlich state},}\ }\href {http://jetpletters.ru/ps/0/article_24247.shtml} {\bibfield  {journal} {\bibinfo  {journal} {JETP Lett.}\ }\textbf {\bibinfo {volume} {28}},\ \bibinfo {pages} {656} (\bibinfo {year} {1978})}\BibitemShut {NoStop}%
\bibitem [{\citenamefont {Brazovskii}\ \emph {et~al.}(1980)\citenamefont {Brazovskii}, \citenamefont {Gordyunin},\ and\ \citenamefont {Kirova}}]{Brazovskii2}%
  \BibitemOpen
  \bibfield  {author} {\bibinfo {author} {\bibfnamefont {S.~A.}\ \bibnamefont {Brazovskii}}, \bibinfo {author} {\bibfnamefont {S.~A.}\ \bibnamefont {Gordyunin}}, \ and\ \bibinfo {author} {\bibfnamefont {N.~N.}\ \bibnamefont {Kirova}},\ }\bibfield  {title} {\enquote {\bibinfo {title} {An exact solution of the peierls model with an arbitrary number of electrons in the unit cell},}\ }\href {http://jetpletters.ru/ps/0/article_20458.shtml} {\bibfield  {journal} {\bibinfo  {journal} {JETP Lett.}\ }\textbf {\bibinfo {volume} {31}},\ \bibinfo {pages} {486} (\bibinfo {year} {1980})}\BibitemShut {NoStop}%
\bibitem [{\citenamefont {Takayama}\ \emph {et~al.}(1980)\citenamefont {Takayama}, \citenamefont {Lin-Liu},\ and\ \citenamefont {Maki}}]{Takayama}%
  \BibitemOpen
  \bibfield  {author} {\bibinfo {author} {\bibfnamefont {Hajime}\ \bibnamefont {Takayama}}, \bibinfo {author} {\bibfnamefont {Y.~R.}\ \bibnamefont {Lin-Liu}}, \ and\ \bibinfo {author} {\bibfnamefont {Kazumi}\ \bibnamefont {Maki}},\ }\bibfield  {title} {\enquote {\bibinfo {title} {Continuum model for solitons in polyacetylene},}\ }\href {\doibase 10.1103/PhysRevB.21.2388} {\bibfield  {journal} {\bibinfo  {journal} {Phys. Rev. B}\ }\textbf {\bibinfo {volume} {21}},\ \bibinfo {pages} {2388--2393} (\bibinfo {year} {1980})}\BibitemShut {NoStop}%
\bibitem [{\citenamefont {Horovitz}(1980)}]{HOROVITZ198061}%
  \BibitemOpen
  \bibfield  {author} {\bibinfo {author} {\bibfnamefont {B.}~\bibnamefont {Horovitz}},\ }\bibfield  {title} {\enquote {\bibinfo {title} {The mixed peierls phase and metallic polyacetylene},}\ }\href {\doibase https://doi.org/10.1016/0038-1098(80)90630-4} {\bibfield  {journal} {\bibinfo  {journal} {Solid State Communications}\ }\textbf {\bibinfo {volume} {34}},\ \bibinfo {pages} {61--64} (\bibinfo {year} {1980})}\BibitemShut {NoStop}%
\bibitem [{\citenamefont {Rice}(1979)}]{RICE1979152}%
  \BibitemOpen
  \bibfield  {author} {\bibinfo {author} {\bibfnamefont {Michael~J.}\ \bibnamefont {Rice}},\ }\bibfield  {title} {\enquote {\bibinfo {title} {Charged $\pi$-phase kinks in lightly doped polyacetylene},}\ }\href {\doibase https://doi.org/10.1016/0375-9601(79)90905-8} {\bibfield  {journal} {\bibinfo  {journal} {Physics Letters A}\ }\textbf {\bibinfo {volume} {71}},\ \bibinfo {pages} {152--154} (\bibinfo {year} {1979})}\BibitemShut {NoStop}%
\bibitem [{\citenamefont {Monceau}\ \emph {et~al.}(2001)\citenamefont {Monceau}, \citenamefont {Nad},\ and\ \citenamefont {Brazovskii}}]{Monceau}%
  \BibitemOpen
  \bibfield  {author} {\bibinfo {author} {\bibfnamefont {P.}~\bibnamefont {Monceau}}, \bibinfo {author} {\bibfnamefont {F.~Ya.}\ \bibnamefont {Nad}}, \ and\ \bibinfo {author} {\bibfnamefont {S.}~\bibnamefont {Brazovskii}},\ }\bibfield  {title} {\enquote {\bibinfo {title} {Ferroelectric mott-hubbard phase of organic $(\mathrm{TMTTF}{)}_{2}\mathit{X}$ conductors},}\ }\href {\doibase 10.1103/PhysRevLett.86.4080} {\bibfield  {journal} {\bibinfo  {journal} {Phys. Rev. Lett.}\ }\textbf {\bibinfo {volume} {86}},\ \bibinfo {pages} {4080--4083} (\bibinfo {year} {2001})}\BibitemShut {NoStop}%
\bibitem [{\citenamefont {Rice}\ and\ \citenamefont {Mele}(1982)}]{meleandrice}%
  \BibitemOpen
  \bibfield  {author} {\bibinfo {author} {\bibfnamefont {M.~J.}\ \bibnamefont {Rice}}\ and\ \bibinfo {author} {\bibfnamefont {E.~J.}\ \bibnamefont {Mele}},\ }\bibfield  {title} {\enquote {\bibinfo {title} {Elementary excitations of a linearly conjugated diatomic polymer},}\ }\href {\doibase 10.1103/PhysRevLett.49.1455} {\bibfield  {journal} {\bibinfo  {journal} {Phys. Rev. Lett.}\ }\textbf {\bibinfo {volume} {49}},\ \bibinfo {pages} {1455--1459} (\bibinfo {year} {1982})}\BibitemShut {NoStop}%
\bibitem [{\citenamefont {Jackiw}\ and\ \citenamefont {Rebbi}(1976)}]{jackiw}%
  \BibitemOpen
  \bibfield  {author} {\bibinfo {author} {\bibfnamefont {R.}~\bibnamefont {Jackiw}}\ and\ \bibinfo {author} {\bibfnamefont {C.}~\bibnamefont {Rebbi}},\ }\bibfield  {title} {\enquote {\bibinfo {title} {Solitons with fermion number \textonehalf{}},}\ }\href {\doibase 10.1103/PhysRevD.13.3398} {\bibfield  {journal} {\bibinfo  {journal} {Phys. Rev. D}\ }\textbf {\bibinfo {volume} {13}},\ \bibinfo {pages} {3398--3409} (\bibinfo {year} {1976})}\BibitemShut {NoStop}%
\bibitem [{\citenamefont {Kivelson}(1983)}]{kivelson}%
  \BibitemOpen
  \bibfield  {author} {\bibinfo {author} {\bibfnamefont {S.}~\bibnamefont {Kivelson}},\ }\bibfield  {title} {\enquote {\bibinfo {title} {Solitons with adjustable charge in a commensurate peierls insulator},}\ }\href {\doibase 10.1103/PhysRevB.28.2653} {\bibfield  {journal} {\bibinfo  {journal} {Phys. Rev. B}\ }\textbf {\bibinfo {volume} {28}},\ \bibinfo {pages} {2653--2658} (\bibinfo {year} {1983})}\BibitemShut {NoStop}%
\bibitem [{\citenamefont {Zhao}\ \emph {et~al.}(2023)\citenamefont {Zhao}, \citenamefont {Han}, \citenamefont {Kivelson},\ and\ \citenamefont {Esterlis}}]{sijia}%
  \BibitemOpen
  \bibfield  {author} {\bibinfo {author} {\bibfnamefont {Sijia}\ \bibnamefont {Zhao}}, \bibinfo {author} {\bibfnamefont {Zhaoyu}\ \bibnamefont {Han}}, \bibinfo {author} {\bibfnamefont {Steven~A.}\ \bibnamefont {Kivelson}}, \ and\ \bibinfo {author} {\bibfnamefont {Ilya}\ \bibnamefont {Esterlis}},\ }\bibfield  {title} {\enquote {\bibinfo {title} {One-dimensional holstein model revisited},}\ }\href {\doibase 10.1103/PhysRevB.107.075142} {\bibfield  {journal} {\bibinfo  {journal} {Phys. Rev. B}\ }\textbf {\bibinfo {volume} {107}},\ \bibinfo {pages} {075142} (\bibinfo {year} {2023})}\BibitemShut {NoStop}%
\bibitem [{Note1()}]{Note1}%
  \BibitemOpen
  \bibinfo {note} {At mean-field level, the transition from the CDW to the BDW always occurs through an intermediate FE phase, but this phase gets exponentially small as quantum fluctuations become increasingly strong. It is an open question whether it becomes a direct, fluctuation-driven first-order transition when effects beyond mean-field theory are included.}\BibitemShut {Stop}%
\bibitem [{\citenamefont {Fradkin}\ and\ \citenamefont {Hirsch}(1983{\natexlab{b}})}]{fradkinhirsch}%
  \BibitemOpen
  \bibfield  {author} {\bibinfo {author} {\bibfnamefont {Eduardo}\ \bibnamefont {Fradkin}}\ and\ \bibinfo {author} {\bibfnamefont {Jorge~E.}\ \bibnamefont {Hirsch}},\ }\bibfield  {title} {\enquote {\bibinfo {title} {Phase diagram of one-dimensional electron-phonon systems. i. the su-schrieffer-heeger model},}\ }\href {\doibase 10.1103/PhysRevB.27.1680} {\bibfield  {journal} {\bibinfo  {journal} {Phys. Rev. B}\ }\textbf {\bibinfo {volume} {27}},\ \bibinfo {pages} {1680--1697} (\bibinfo {year} {1983}{\natexlab{b}})}\BibitemShut {NoStop}%
\bibitem [{\citenamefont {Cross}\ and\ \citenamefont {Fisher}(1979)}]{crossandfisher}%
  \BibitemOpen
  \bibfield  {author} {\bibinfo {author} {\bibfnamefont {M.~C.}\ \bibnamefont {Cross}}\ and\ \bibinfo {author} {\bibfnamefont {Daniel~S.}\ \bibnamefont {Fisher}},\ }\bibfield  {title} {\enquote {\bibinfo {title} {A new theory of the spin-peierls transition with special relevance to the experiments on ttfcubdt},}\ }\href {\doibase 10.1103/PhysRevB.19.402} {\bibfield  {journal} {\bibinfo  {journal} {Phys. Rev. B}\ }\textbf {\bibinfo {volume} {19}},\ \bibinfo {pages} {402--419} (\bibinfo {year} {1979})}\BibitemShut {NoStop}%
\bibitem [{\citenamefont {Heeger}\ \emph {et~al.}(1988)\citenamefont {Heeger}, \citenamefont {Kivelson}, \citenamefont {Schrieffer},\ and\ \citenamefont {Su}}]{HKSS}%
  \BibitemOpen
  \bibfield  {author} {\bibinfo {author} {\bibfnamefont {A.~J.}\ \bibnamefont {Heeger}}, \bibinfo {author} {\bibfnamefont {S.}~\bibnamefont {Kivelson}}, \bibinfo {author} {\bibfnamefont {J.~R.}\ \bibnamefont {Schrieffer}}, \ and\ \bibinfo {author} {\bibfnamefont {W.~P.}\ \bibnamefont {Su}},\ }\bibfield  {title} {\enquote {\bibinfo {title} {Solitons in conducting polymers},}\ }\href {\doibase 10.1103/RevModPhys.60.781} {\bibfield  {journal} {\bibinfo  {journal} {Rev. Mod. Phys.}\ }\textbf {\bibinfo {volume} {60}},\ \bibinfo {pages} {781--850} (\bibinfo {year} {1988})}\BibitemShut {NoStop}%
\bibitem [{\citenamefont {Kivelson}\ and\ \citenamefont {Heim}(1982)}]{kivelsonheim}%
  \BibitemOpen
  \bibfield  {author} {\bibinfo {author} {\bibfnamefont {S.}~\bibnamefont {Kivelson}}\ and\ \bibinfo {author} {\bibfnamefont {D.~E.}\ \bibnamefont {Heim}},\ }\bibfield  {title} {\enquote {\bibinfo {title} {Hubbard versus peierls and the su-schrieffer-heeger model of polyacetylene},}\ }\href {\doibase 10.1103/PhysRevB.26.4278} {\bibfield  {journal} {\bibinfo  {journal} {Phys. Rev. B}\ }\textbf {\bibinfo {volume} {26}},\ \bibinfo {pages} {4278--4292} (\bibinfo {year} {1982})}\BibitemShut {NoStop}%
\bibitem [{\citenamefont {Su}\ and\ \citenamefont {Schrieffer}(1980)}]{suschrieffer}%
  \BibitemOpen
  \bibfield  {author} {\bibinfo {author} {\bibfnamefont {W.~P.}\ \bibnamefont {Su}}\ and\ \bibinfo {author} {\bibfnamefont {J.~R.}\ \bibnamefont {Schrieffer}},\ }\bibfield  {title} {\enquote {\bibinfo {title} {Soliton dynamics in polyacetylene},}\ }\href {\doibase 10.1073/pnas.77.10.5626} {\bibfield  {journal} {\bibinfo  {journal} {Proceedings of the National Academy of Sciences}\ }\textbf {\bibinfo {volume} {77}},\ \bibinfo {pages} {5626--5629} (\bibinfo {year} {1980})},\ \Eprint {http://arxiv.org/abs/https://www.pnas.org/doi/pdf/10.1073/pnas.77.10.5626} {https://www.pnas.org/doi/pdf/10.1073/pnas.77.10.5626} \BibitemShut {NoStop}%
\bibitem [{\citenamefont {Goldstone}\ and\ \citenamefont {Wilczek}(1981)}]{goldstonewilczek}%
  \BibitemOpen
  \bibfield  {author} {\bibinfo {author} {\bibfnamefont {Jeffrey}\ \bibnamefont {Goldstone}}\ and\ \bibinfo {author} {\bibfnamefont {Frank}\ \bibnamefont {Wilczek}},\ }\bibfield  {title} {\enquote {\bibinfo {title} {Fractional quantum numbers on solitons},}\ }\href {\doibase 10.1103/PhysRevLett.47.986} {\bibfield  {journal} {\bibinfo  {journal} {Phys. Rev. Lett.}\ }\textbf {\bibinfo {volume} {47}},\ \bibinfo {pages} {986--989} (\bibinfo {year} {1981})}\BibitemShut {NoStop}%
\bibitem [{S0()}]{S0}%
  \BibitemOpen
  \href@noop {} {\ }\bibinfo {note} {See also Supplemental Material Section B for more details.}\BibitemShut {Stop}%
\bibitem [{Note2()}]{Note2}%
  \BibitemOpen
  \bibinfo {note} {In the case of the neutral composite with the quantum numbers of $S_0$, it is likely that the lowest energy saddle point configuration is actually an amplitude mode, with $\theta _j=$ constant and $A_j$ changing sign across the width of the soliton.}\BibitemShut {Stop}%
\bibitem [{\citenamefont {Su}(1981)}]{WPSUpaper}%
  \BibitemOpen
  \bibfield  {author} {\bibinfo {author} {\bibfnamefont {W.~P.}\ \bibnamefont {Su}},\ }\bibfield  {title} {\enquote {\bibinfo {title} {Real time dynamics in polyacetylene},}\ }\href {\doibase 10.1080/00268948108075246} {\bibfield  {journal} {\bibinfo  {journal} {Molecular Crystals and Liquid Crystals}\ }\textbf {\bibinfo {volume} {77}},\ \bibinfo {pages} {265--275} (\bibinfo {year} {1981})}\BibitemShut {NoStop}%
\bibitem [{\citenamefont {Fishman}\ \emph {et~al.}(2022)\citenamefont {Fishman}, \citenamefont {White},\ and\ \citenamefont {Stoudenmire}}]{itensor}%
  \BibitemOpen
  \bibfield  {author} {\bibinfo {author} {\bibfnamefont {Matthew}\ \bibnamefont {Fishman}}, \bibinfo {author} {\bibfnamefont {Steven~R.}\ \bibnamefont {White}}, \ and\ \bibinfo {author} {\bibfnamefont {E.~Miles}\ \bibnamefont {Stoudenmire}},\ }\bibfield  {title} {\enquote {\bibinfo {title} {The itensor software library for tensor network calculations},}\ }\href {\doibase 10.21468/SciPostPhysCodeb.4} {\bibfield  {journal} {\bibinfo  {journal} {SciPost Phys. Codebases}\ ,\ \bibinfo {pages} {4}} (\bibinfo {year} {2022})}\BibitemShut {NoStop}%
\end{thebibliography}%

\clearpage
\renewcommand{\thefigure}{S\arabic{figure}}
\setcounter{figure}{0}
\onecolumngrid
\subsection*{\large{Supplemental Material}}
\subsection*{A. Soliton configurations}
Here we summarize the results of additional DMRG calculations concerning a variety of soliton containing ground-states.  As summarized in TABLE~\ref{Table1}, various situations with different numbers and types of solitons can be produced in response to mismatched boundary conditions, dopings, unequal number of spin up and spin down electrons, or combinations of these three conditions.

\begin{figure*}[h!]
    \centering
  \includegraphics[width=1.0\linewidth]{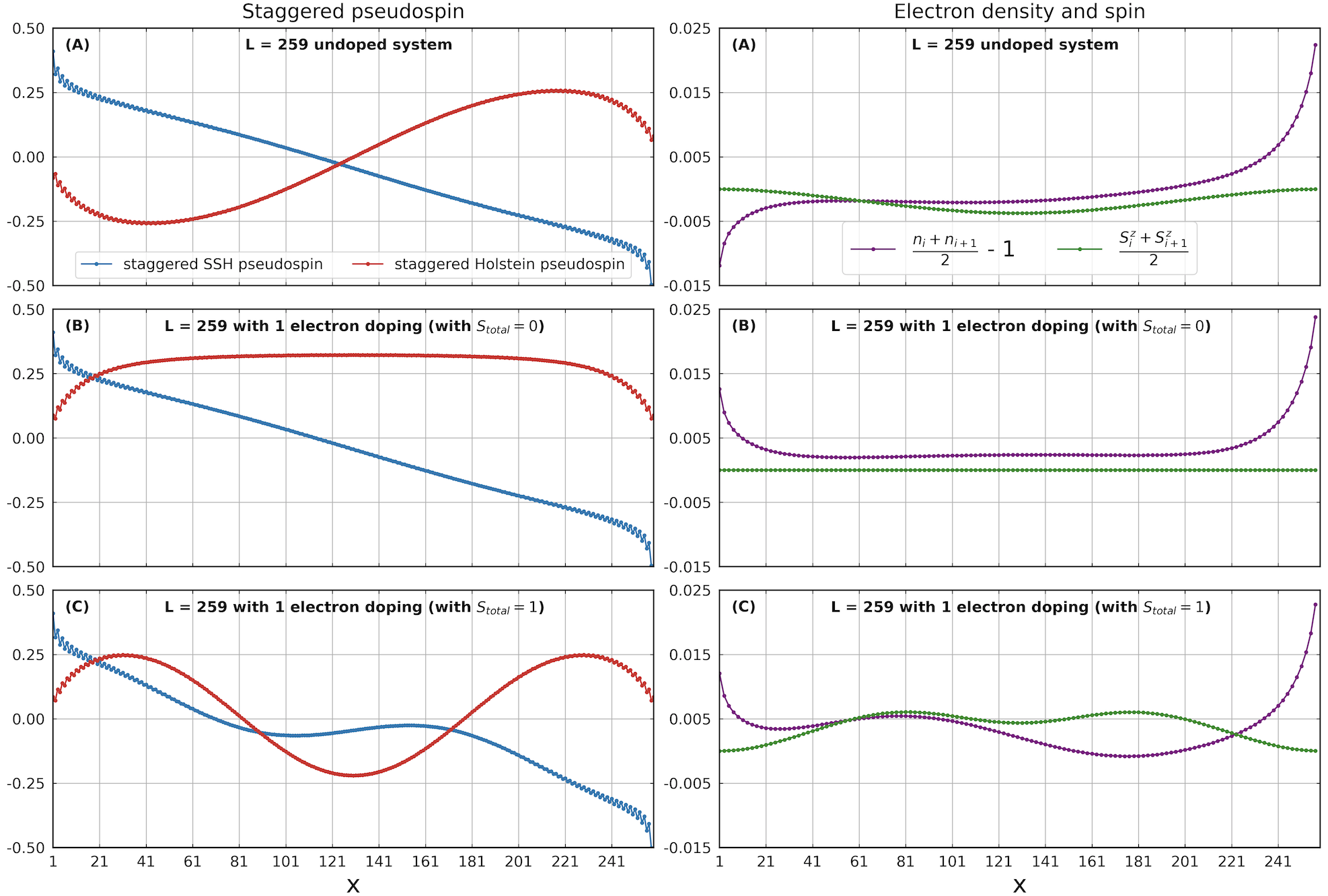}
\caption{\label{fig:S-1}\textbf{Odd-length systems} with (A) no doping (B) doping 1 electron (with $S_{\text{tot}}$=0) and (C) doping 1 electron (with $S_{\text{tot}}$=1). Parameters: $L=259$, $\alpha=0.458$, $\gamma=0.876$, $b_x=h_x=0.14$.}
\end{figure*}

\begin{figure*}[h!]
    \centering
  \includegraphics[width=1.0\linewidth]{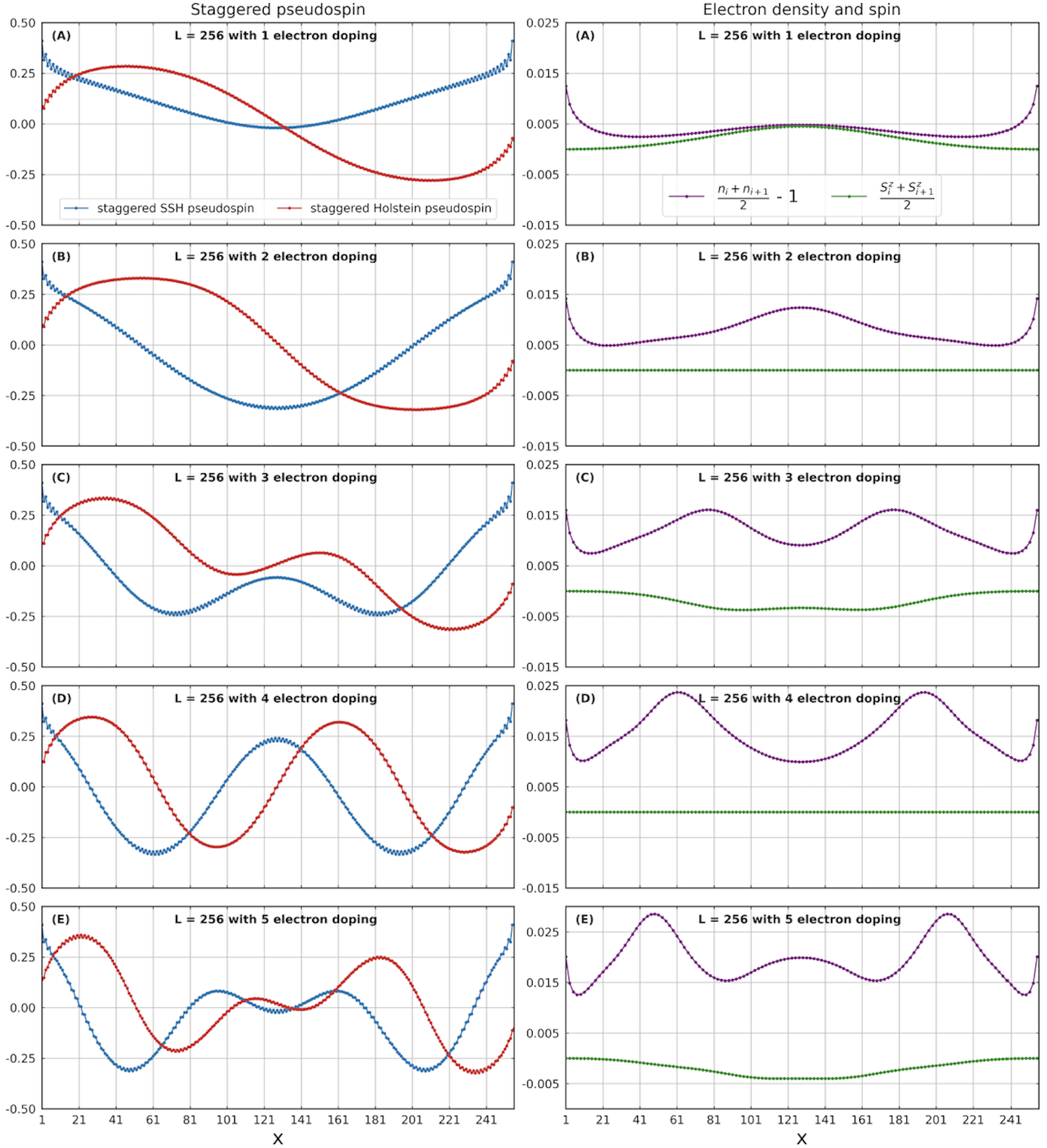}
\caption{\label{fig:S-2}\textbf{Even-length systems with various doping levels.} Parameters: $L=256$, $\alpha=0.458$, $\gamma=0.876$, $b_x=h_x=0.14$.}
\end{figure*}

\begin{figure*}[h!]
    \centering
  \includegraphics[width=1.0\linewidth]{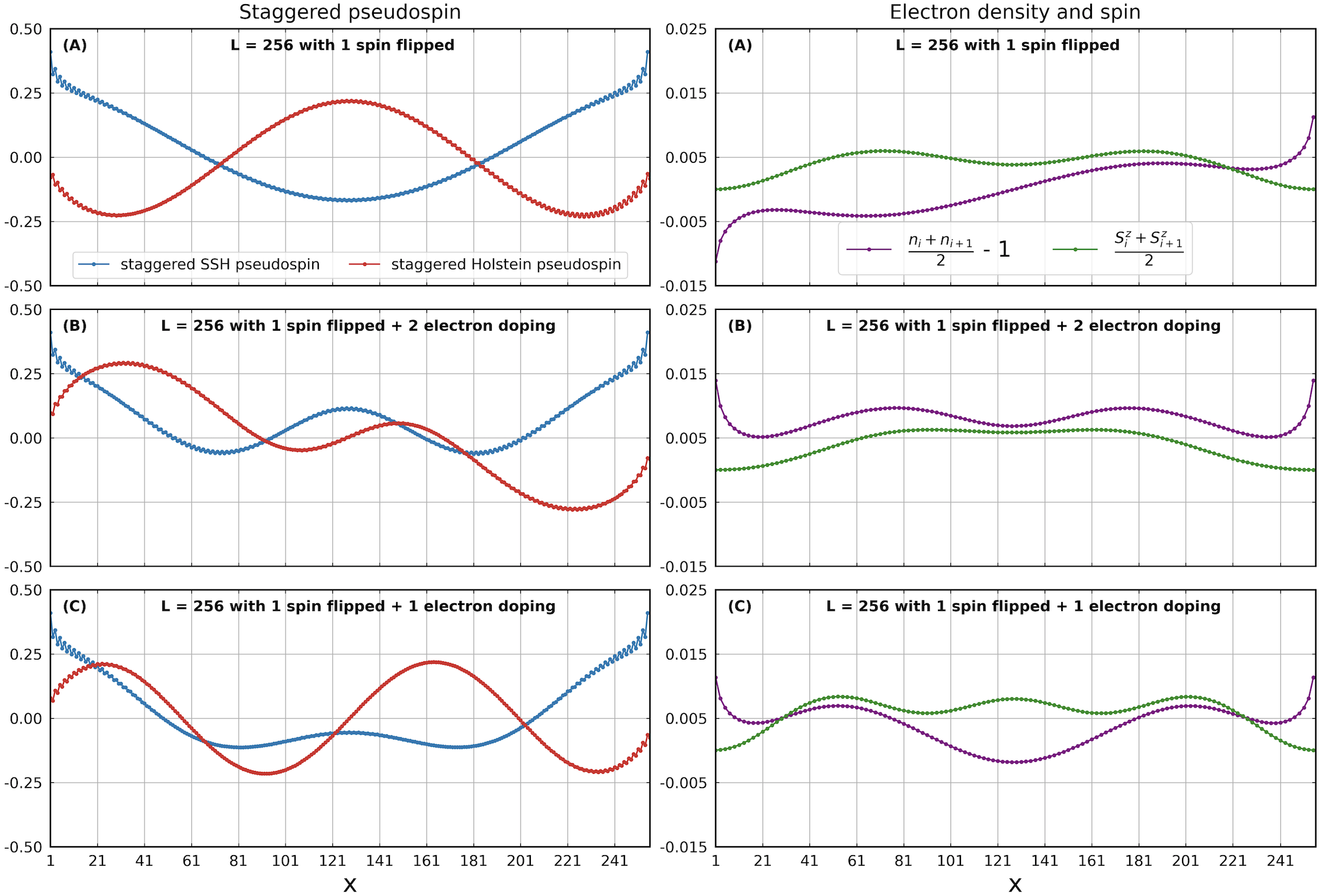}
\caption{\label{fig:S-3}\textbf{Even-length systems with both spin-flipping and doping.} Parameters: $L=256$, $\alpha=0.458$, $\gamma=0.876$, $b_x=h_x=0.14$.}
\end{figure*}

\begin{figure*}[h!]
    \centering
  \includegraphics[width=1.0\linewidth]{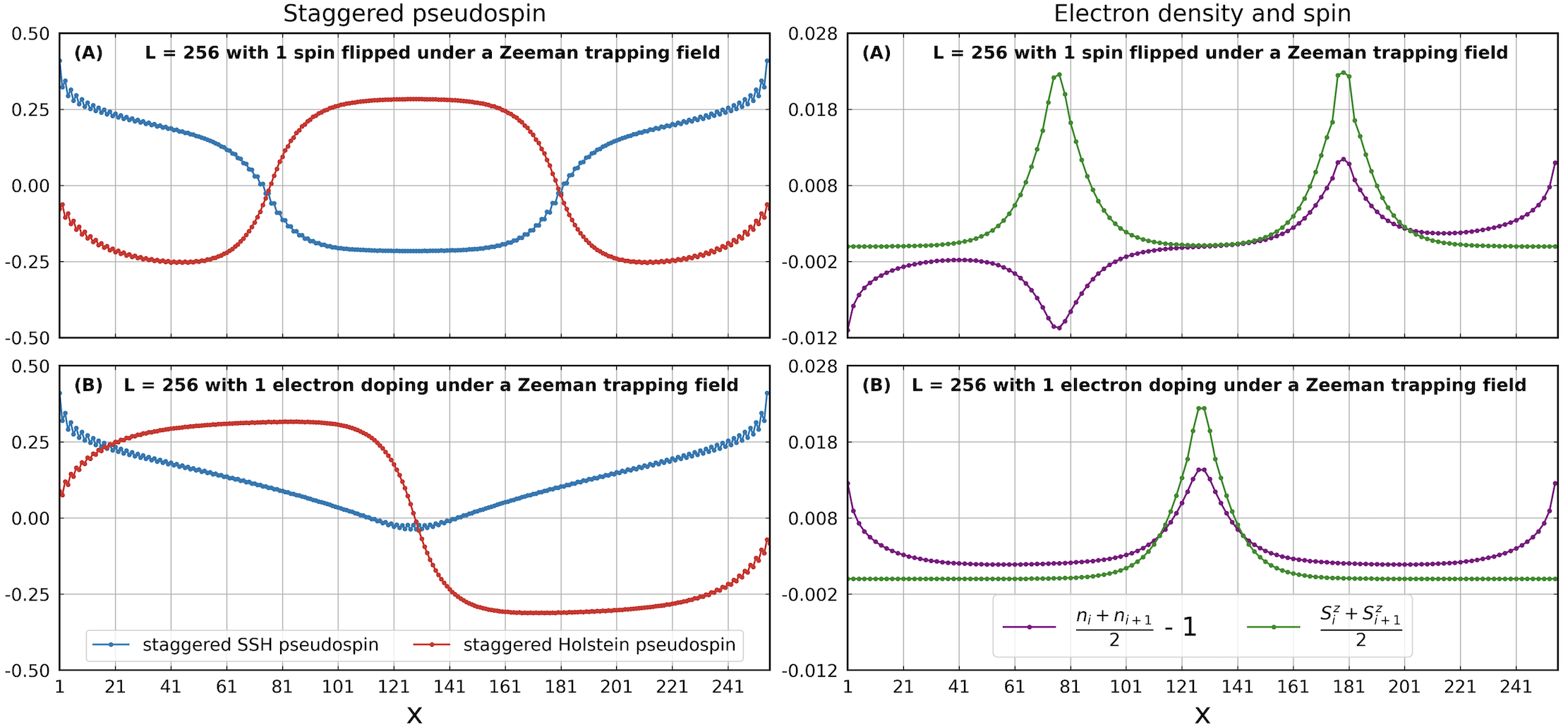}
\caption{\label{fig:S-4}\textbf{Adding a Zeeman trapping field $\bf{H_{\text{trap}}=-0.05S_z}$ on electron spins to trap solitons} for (A) even-length system with one spin-flipped case and (B) even-length system with one-electron doping case as examples. Parameters: $L=256$, $\alpha=0.458$, $\gamma=0.876$, $b_x=h_x=0.14$.}
\end{figure*}

\begin{enumerate}
\item \begin{enumerate}
      \item In an odd-length system with no doping (neutral), a pair of SSH soliton ($S_{\tau}$) and Holstein soliton ($S_{\sigma}$) is generated as shown in Fig.~\ref{fig:S-1} (A).
      
      \item Then if we add one-electron (hole) doping and keep total spin $S_{\text{tot}}=0$, only a single SSH soliton ($S_{\tau}$) remains as shown in Fig.~\ref{fig:oddL-main} in the main text also Fig.~\ref{fig:S-1} (B) here.
      
      \item Similarly, if we add one-electron (hole) doping but make $S_{tot}=1$, a pair of $S_{\tau}$ and $S_{\sigma}$ are now generated on each side together with another $S_{\tau}$ at the center of the system (Fig.~\ref{fig:S-1} (C)).
  \end{enumerate}
\item \begin{enumerate}
      \item In an even-length system with one-electron (hole) doping, an SSH polaron together with a Holstein soliton ($S_{\sigma}$) are produced as mentioned in the main text and shown in Fig.~\ref{fig:S-2} (A). We have also applied a Zeeman trapping field to localize them as shown in Fig.~\ref{fig:S-4} (B).
      
      \item If we add one more electron (hole), then with two-electron (hole) doping we see the SSH polaron disappeared; instead, a pair of SSH solitons ($S_{\tau}$) has now been generated (Fig.~\ref{fig:S-2} (B)). 
      
      \item When we add three electrons (holes), apart from the pair of SSH solitons ($S_{\tau}$), an SSH polaron appears again, and two more $S_{\sigma}$ also shown in the Holstein pseudo-spin configuration (Fig.~\ref{fig:S-2} (C)).
      
      \item When four electrons (holes) are doped, similar to case 2(b), now we find four SSH solitons ($S_{\tau}$) and three Holstein solitons ($S_{\sigma}$) (Fig.~\ref{fig:S-2} (D)), from which we can predict that with six-electron (hole) doping, there will likely be six SSH solitons ($S_{\tau}$) and five Holstein solitons ($S_{\sigma}$).
      \item In the final pure doping case with five-electron (hole) doping, as shown in Fig.~\ref{fig:S-2} (E) an additional pair of $S_{\tau}$ and $S_{\sigma}$ each are generated based on case 2(c), exactly as in the transition from case 2(a) to case 2(c), i.e. from one-electron (hole) doping to three-electron (hole) doping.  
      \item[*] From this series of doping conditions, we discovered a pattern: (with m odd)
      \begin{align}
      \label{eq:modd}
      \text{$m$ - electron doping} &~\Rightarrow~ \text{1 SSH polaron + $(m-1)~S_{\tau}$ + $m~ S_{\sigma}$}\\[1.5ex]
      \text{$(m+1)$ - electron doping}  &~\Rightarrow~ \text{0 SSH polaron + $(m+1)~S_{\tau}$ + $m~S_{\sigma}$}
      \end{align}
  \end{enumerate}
\item \begin{enumerate}
      \item In an even-length system with no dopings but 1 spin flipped, a pair of $S_{\tau}$ and a pair of $S_{\sigma}$ are generated simultaneously (Fig.~\ref{fig:S-3} (A)), which can also be well localized by a Zeeman trapping field on each side as shown in Fig.~\ref{fig:S-4} (A).
      
      \item In pure doping cases, when the total charge $Q_{\text{tot}}$ is changed by 2 with $S_{\text{tot}}=\frac{1}{2}$ fixed, e.g.,  from 2(a) to 2(c), a pair each of $S_{\tau}$ and $S_{\sigma}$ is produced. Now, as a comparison, we consider changing $Q_{\text{tot}}$ by 2 with $S_{\text{tot}}=1$. As shown in Fig.~\ref{fig:S-3} (A) and (B), instead of a pair of $S_{\sigma}$, only a single $S_{\sigma}$ is produced at the center in this case. Likely, this is because the cases with $S_{\text{tot}}=\frac{1}{2}$ have $m$ electron doping with $m$ being an odd number. As summarized in Eq.~\ref{eq:modd}, this will generate an odd number of $S_{\sigma}$, implying a soliton configuration with one $S_{\sigma}$ at the center and $(m-1)/2$ $S_{\sigma}$ on each side. In contrast, in case 3(a) with $S_{\text{tot}}=1$, there are an even number of $S_{\sigma}$ to begin with, i.e., no soliton at the center. So when we increase $Q_{\text{tot}}$ by 2 in case 3(b), the center position becomes the plausible ``first choice'' for additionally generated solitons. 
      
      \item In all of the above cases, $S_{\text{tot}}$ is either 0, $\frac{1}{2}$ or 1. To investigate the differences when $S_{\text{tot}}=\frac{3}{2}$, we consider a scenario with one spin flipped together with one-electron doping ($Q_{\text{tot}}=1, S_{\text{tot}}=\frac{3}{2}$) as shown in Fig.~\ref{fig:S-3} (C). Here, we observe a similar soliton configuration as in case 2(c) with three-electron doping (both have 1 SSH polaron, 2 SSH soliton $S_{\tau}$ and 3 Holstein soliton $S_{\sigma}$, but with different pseudo-spin magnitudes).

      \item[*] Up to different magnitudes, we find 2(c) \& 3(c) have the same number of $S_{\tau}$ and $S_{\sigma}$, and 2(d) \& 3(b) do as well. In both pairs of cases, 
      \begin{align}
      &\Delta Q_{\text{tot}}=Q_{\text{tot}}\big[3(c)\big]-Q_{\text{tot}}\big[2(c)\big]=Q_{\text{tot}}\big[3(b)\big]-Q_{\text{tot}}\big[2(d)\big]=-2\\[1.5ex]
      &\Delta S_{\text{tot}}=S_{\text{tot}}\big[3(c)\big]-S_{\text{tot}}\big[2(c)\big]=S_{\text{tot}}\big[3(b)\big]-S_{\text{tot}}\big[2(d)\big]=1
      \end{align}
      this implies $|Q_{\text{tot}}|=2$ and $|S_{\text{tot}}|=1$ may have the same effect on soliton production, thus they effectively cancel each other out in the above pairs of cases. Surprisingly, this can indeed be confirmed from above: 
      \begin{align}
      &\text{From 2(b) to 2(d):~~~}\Delta Q_{\text{tot}}=2,~\Delta S_{\text{tot}}=0 ~\Rightarrow~ \text{~two more~}S_{\tau~~}\& \text{~~two more~} S_{\sigma}\\[1.5ex]
      &\text{From 1(b) to 1(c):~~~}\Delta Q_{\text{tot}}=0,~\Delta S_{\text{tot}}=1 ~\Rightarrow~ \text{~two more~}S_{\tau~~}\& \text{~~two more~} S_{\sigma}
      \end{align}
      Therefore, we see there is indeed an equivalency between changes in $Q_{\text{tot}}$ and $S_{\text{tot}}$, which is consistent across all cases we have studied.   
  \end{enumerate}
\end{enumerate}

\begin{table}[h!]
    \centering
\includegraphics[width=0.9\linewidth]{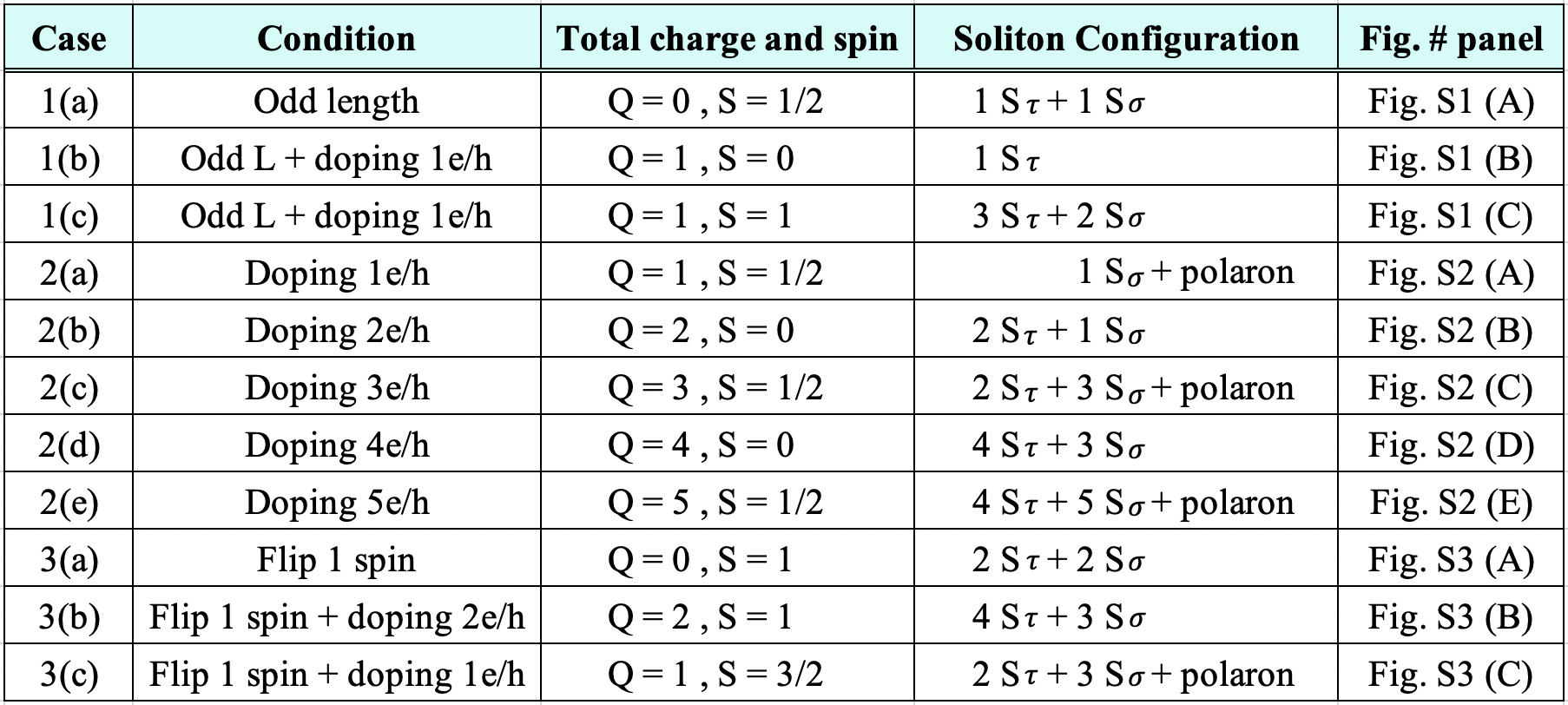}
\caption{\label{Table1}}
\end{table}

\clearpage
\subsection*{B. Exact soliton excitations in the classical limit}
\noindent At $b_x=h_x=0$ limit, we not only know that the ground state is ferroelectric, but we can also construct the exact soliton excitations in terms of the pseudo-spins. Here we use $\tau^z_j = (-1)^j \ \eta(R-j)$ and $\sigma^z_j=(-1)^j$ with $R=30$ as an example in a $L=64$ system. The non-interacting Hamiltonian now is
\begin{align}
\label{eq:H}
H=&-\sum_j \Bigg[t\hat{B}_j +\alpha(-1)^j\eta(R-j)\left(\hat B_j-B\right)\Bigg]\nonumber -\sum_j \Bigg[\gamma(-1)^j\left(\hat n_j-n\right)\Bigg]
\end{align}
Then through exact diagonalization, we can get the electron density profile of the soliton and anti-soliton states as shown in the right panel of Fig.~\ref{fig:theta} in the main text, where for a set of strong couplings, the soliton width is much shorter than that in the weak-coupling case as expected.

\subsection*{C. Mean-field calculation}
\noindent To obtain a mean-field phase diagram, we introduce the trial Hamiltonian: 
\begin{equation}
\begin{split}
H_{\text{tr}}&=-\sum_j \Big[\Big(1+\alpha\tau(-1)^j\Big)\left(\hat B_j-B\right)+\alpha m(-1)^j\tau^z_j+b_x\tau^x_j\Big]-\sum_j\Big[\gamma\sigma(-1)^j\left(\hat n_j-n\right) + \gamma M(-1)^j\sigma^z_j+h_x\sigma^x_j\Big]
\end{split}   
\end{equation}
With some standard algebra we can show that at zero temperature the variational energy is equal to:
\begin{equation}
\begin{split}
F_{\text{var}}&=\langle H \rangle_{\text{tr}}= - \sum_{j}  \Big[ \langle \hat B_j\rangle_{\text{tr}}+ \alpha \langle\tau_j^z\rangle_{\text{tr}}\langle\hat B_j-B\rangle_{\text{tr}} + b_x\langle\tau_j^x\rangle_{\text{tr}}\Big]
- \sum_j \Big[ \gamma \langle\sigma_j^z\rangle_{\text{tr}}\langle\hat n_j-n\rangle_{\text{tr}} + h_x\langle\sigma_j^x\rangle_{\text{tr}}\Big] \\[2ex]
=&-N\Big(\frac{b^2_x}{2E}+\frac{h^2_x}{2\lambda}\Big)-\sum_j \Big((-1)^j+\alpha \frac{m}{2E}\Big)\Big(\frac{4\alpha\tau}{\pi\sqrt{1+(\gamma\sigma)^2/4}}\Big)\Big(\frac{1+\frac{(\gamma\sigma)^2}{4}}{1-\alpha^2\tau^2}\Big)\Bigg[\textbf{EI}_1\bigg(\sqrt{\frac{1-\alpha^2\tau^2}{1+(\gamma\sigma)^2/4}}~\bigg)-\textbf{EI}_2\bigg(\sqrt{\frac{1-\alpha^2\tau^2}{1+(\gamma\sigma)^2/4}}~\bigg)\Bigg]\\[2ex]
&-\sum_j \gamma\frac{M}{2\lambda}\Big(\frac{\gamma\sigma}{\pi\sqrt{1+(\gamma\sigma)^2/4}}\Big)\textbf{EI}_1\bigg(\sqrt{\frac{1-\alpha^2\tau^2}{1+(\gamma\sigma)^2/4}}~\bigg)
\end{split}
\end{equation}
where $\textbf{EI}_1(k) = \int^{\frac{\pi}{2}}_0\frac{d\theta}{\sqrt{1-k^2\sin^2\theta}}$ and $\textbf{EI}_2(k) = \int^{\frac{\pi}{2}}_0\sqrt{1-k^2\sin^2\theta}\ d\theta$ are the complete elliptic integral of the first and second kind. Then by minimizing $F_{\text{var}}$ we get four self-consistency relations as mentioned in the main text:
\begin{equation}
\begin{split}
m&=\frac{4\alpha^2\tau\sqrt{1+(\gamma\sigma)^2/4}}{\pi(1-\alpha^2\tau^2)}\Bigg[\textbf{EI}_1\Big(\frac{1-\alpha^2\tau^2}{1+(\gamma\sigma)^2/4}\Big)-\textbf{EI}_2\Big(\frac{1-\alpha^2\tau^2}{1+(\gamma\sigma)^2/4}\Big)\Bigg]\\[2ex]
M&=\Big(\frac{\gamma^2\sigma}{\pi\sqrt{1+(\gamma\sigma)^2/4}}\Big)\textbf{EI}_1\Big(\frac{1-\alpha^2\tau^2}{1+(\gamma\sigma)^2/4}\Big)\\[2ex]
\tau&=\frac{m}{2\sqrt{m^2+b^2_x}}=f(m)\\[2ex]
\sigma&=\frac{M}{2\sqrt{M^2+h^2_x}}=g(M)
\end{split}
\end{equation}

\noindent By expressing $\tau$ and $\sigma$ as functions of m and M, we can recast $F_{\text{var}}$ to be a function of m and M only:

\begin{equation}
\begin{split}
F_{\text{var}}(m,M)/N=&-\Big(\frac{b^2_x}{2\sqrt{m^2+b^2_x}}+\frac{h^2_x}{2\sqrt{M^2+h^2_x}}\Big)-\frac{4\alpha^2f^2(m)\sqrt{1+\frac{\gamma^2}{4}g^2(M)}}{\pi(1-\alpha^2f^2(m))}\Bigg[\textbf{EI}_1\Big(\frac{1-\alpha^2 f^2(m)}{1+\frac{\gamma^2}{4}g^2(M)}\Big)-\textbf{EI}_2\Big(\frac{1-\alpha^2 f^2(m)}{1+\frac{\gamma^2}{4}g^2(M)}\Big)\Bigg]\\[2ex]
&-\frac{\gamma^2 g^2(M)}{\pi\sqrt{1+\frac{\gamma^2}{4}g^2(M)}}\textbf{EI}_1\Big(\frac{1-\alpha^2 f^2(m)}{1+\frac{\gamma^2}{4}g^2(M)}\Big)\\[2ex]
\end{split}
\end{equation}
then at any chosen values of couplings, we can minimize it to get m and M and thus the phase boundary between the ferroelectric phase and the pure CDW or BDW phases.

\end{document}